\documentclass[aip,jcp,amsmath,amssymb,reprint,superscriptaddress]{revtex4-2}

\usepackage[dvipdfmx]{graphicx}
\usepackage{epsf}
\usepackage{bm}
\usepackage{lipsum}
\usepackage[normalem]{ulem}
\usepackage{txfonts}
\usepackage{color}

\usepackage{physics}
\usepackage{booktabs}
\usepackage{multirow}
\newcommand{\ob}{$\mathrm{O}^1$}
\newcommand{\os}{$\mathrm{O}^2$}
\newcommand{\om}{$\mathrm{O}^3$}
\newcommand{\ow}{$\mathrm{O}^\mathrm{w}$}

\newcommand{\D}{\Delta}

\newcommand{\mr}{\mathrm}
\newcommand{\hb}{_\mathrm{HB}}

\newcommand{\phb}{P_\mathrm{HB}}
\newcommand{\thb}{\tau_\mathrm{HB}}
\newcommand{\pbw}{P_\mathrm{BW}}

\newcommand{\eq}[1]{Eq.~(\ref{eq:#1})}
\newcommand{\fig}[1]{Fig.~\ref{fig:#1}}
\newcommand{\tab}[1]{Table.~\ref{tab:#1}}

\begin{document}
\title{Revealing the hidden dynamics of confined water in acrylate
polymers: Insights from hydrogen-bond lifetime analysis}

\author{Kokoro Shikata}
\affiliation{Division of Chemical Engineering, Department of Materials Engineering Science, Graduate School of Engineering Science, Osaka University, Toyonaka, Osaka 560-8531, Japan}

\author{Takuma Kikutsuji}
\affiliation{Division of Chemical Engineering, Department of Materials Engineering Science, Graduate School of Engineering Science, Osaka University, Toyonaka, Osaka 560-8531, Japan}

\author{Nobuhiro Yasoshima}
\affiliation{Division of Chemical Engineering, Department of Materials Engineering Science, Graduate School of Engineering Science, Osaka University, Toyonaka, Osaka 560-8531, Japan}
\affiliation{Department of Information and Computer Engineering, National
Institute of Technology, Toyota College, 2-1 Eiseicho, Toyota, Aichi,
471-8525, Japan}

\author{Kang Kim}
\email{kk@cheng.es.osaka-u.ac.jp}
\affiliation{Division of Chemical Engineering, Department of Materials Engineering Science, Graduate School of Engineering Science, Osaka University, Toyonaka, Osaka 560-8531, Japan}

\author{Nobuyuki Matubayasi}
\email{nobuyuki@cheng.es.osaka-u.ac.jp}
\affiliation{Division of Chemical Engineering, Department of Materials Engineering Science, Graduate School of Engineering Science, Osaka University, Toyonaka, Osaka 560-8531, Japan}

\date{\today}

\begin{abstract}
Polymers contain functional groups that participate in hydrogen
 bond (H-bond) with water molecules, establishing a robust 
 H-bond network that influences bulk properties.
 This study
 utilized molecular dynamics (MD) simulations to examine the
 H-bonding dynamics of water molecules confined within three
 poly(meth)acrylates: poly(2-methoxyethyl acrylate) (PMEA),
 poly(2-hydroxyethyl methacrylate) (PHEMA), and poly(1-methoxymethyl
 acrylate) (PMC1A). 
Results showed that H-bonding dynamics significantly
 slowed as the water content decreased. 
Additionally, the diffusion of
 water molecules and its correlation with H-bond breakage were
 analyzed.
Our findings suggest that when the H-bonds between water molecules and
 the methoxy oxygen of PMEA are disrupted, those water
 molecules persist in close proximity and do not diffuse on a picosecond timescale. 
In contrast, the water molecules H-bonded with the
 hydroxy oxygen of PHEMA and the methoxy oxygen of PMC1A diffuse
 concomitantly with the breakage of H-bonds.
These results provide an in-depth understanding of the
 impact of polymer functional groups on H-bonding dynamics.
\end{abstract}
\maketitle

\section{Introduction}

Poly(2-methoxyethyl acrylate) (PMEA) has gathered significant attention
as a polymeric material with a high degree of blood
compatibility.~\cite{tanaka2000Blood, tanaka2002Study} 
It is hypothesized that the formation of a hydration layer on the
polymer surface may be the primary contributor to its blood
compatibility.
In particular, the presence of loosely interacting water molecules on its surface is
thought to play a crucial role in its ability to inhibit protein
adsorption and denaturation in the event of contact with blood, which is
a well-known precipitant of thrombus formation.~\cite{tanaka2004Effect}
The molecular-level insights of the interaction and dynamics of water at
the surface of PMEA are thus sought to understand and improve the
functions of the polymer surfaces.
Specifically, accurate characterization of the motional time scale of water molecules confined
within a polymer matrix is crucial since these water
molecules in the proximity of the polymer surface are known to play a
significant role in the underlying mechanism of blood compatibility.

Experimentally, differential scanning calorimetry (DSC) measurements revealed the
presence of three distinct states of water in PMEA: free water, which
freezes at 0$^\circ$C; intermediate water, which
crystallizes near -40$^\circ$C during the temperature increase process; and
non-freezing water, which does not freeze even at -100$^\circ$C.~\cite{tanaka2000Cold}
Furthermore, the hydration
state of water molecules in PMEA was investigated through 
infrared (IR) spectroscopy.~\cite{kitano2001Structure, kitano2005Correlation, morita2007TimeResolved} 
The H-bonding bands of the carbonyl group in PMEA were detected,
signifying the presence of H-bonded non-freezing water molecules. 
The
water molecules that interact with the oxygen of the methoxy group in
PMEA are believed to exhibit an anomalous mobility, which is
characterized as intermediate water. 
In addition, the detection of water molecules
possessing a H-bonding structure similar to that of bulk water implies
the existence of free water. 
However, capturing more precise 
pictures of water molecules confined within polymers remains challenging,
since 
the dynamical states of water are diverse as classified into
non-freezing water ($10^{-8} - 10^{-6}$ s), intermediate
water ($10^{-10} - 10^{-9}$ s), and free water ($10^{-12} - 10^{-11}$
s), with their respective timescales determined through nuclear magnetic resonance (NMR)
measurements.~\cite{miwa2009Network, tsuruta2010Role}

Molecular dynamics (MD) simulations serve as a formidable method to
achieve an in-depth comprehension of the structural and dynamic 
features of water molecules confined within PMEA.
To this end, various MD simulations of hydrated PMEA have been
conducted.~\cite{nagumo2013Computational, nagumo2019Molecular,
nagumo2021Interactionsa, yasoshima2017DiffusionControlled,
kishinaka2019Molecular, yasoshima2021Molecular, yasoshima2022Adsorption}
A particular focus is to investigate the molecular structure and
vibrational spectra of water molecules at the water/polymer interface~\cite{kishinaka2019Molecular}
and within hydrated PMEA.~\cite{yasoshima2021Molecular}
In studies by Kuo \textit{et al.}, MD simulations were utilized to
classify water molecules confined within PMEA based on their H-bonding
interactions with polymer oxygen.~\cite{kuo2019Analyses,
kuo2020Elucidating, kuo2020Molecular, kuo2021Effects} 
It was proposed that 
non-bound water (NBW) denotes water
molecules devoid of H-bonds with polymer oxygen, while those
exhibiting a single H-bond were denoted as one-bound water (BW1), and
those displaying a double H-bond were denoted as two-bound water
(BW2).
Additionally, the proportion of BW2, BW1, and NBW was found to exhibit a
similar dependence on water content as the proportion of non-freezing
water, intermediate water, and free water, as determined through
DSC analysis.

\begin{figure*}[t]
\centering
\includegraphics[width=0.8\textwidth]{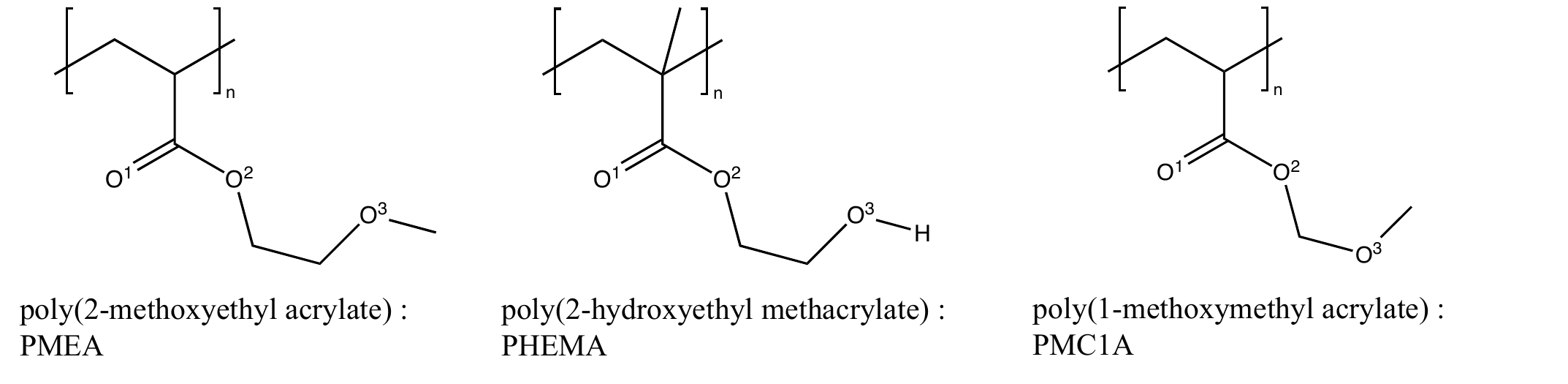}
\caption{Structures of the polymers studied in this paper.}
\label{fig:polymer}
\end{figure*}

\begin{figure}[t]
\centering
\includegraphics[width=1.0\linewidth]{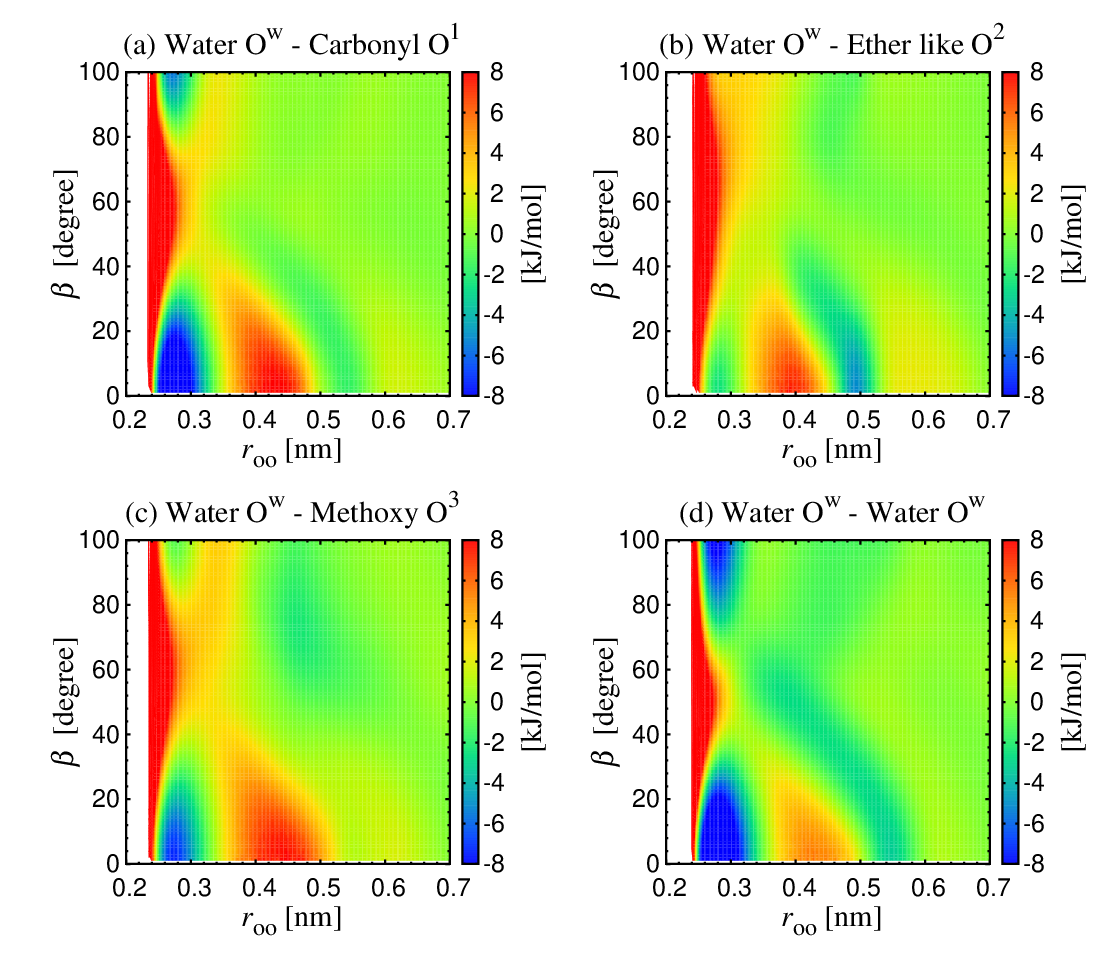}
\caption{2D PMF $W(r_\mr{oo},\beta)$ between water oxygen ({\ow})
 and acceptor oxygen [(a) carbonyl {\ob}, (b) ether-like {\os}, (c) methoxy {\om},
 and (d) water {\ow}] in PMEA-water system
 at 9 wt\%.}
\label{fig:pmf_pmea}
\end{figure}

\begin{figure}[t]
\centering
\includegraphics[width=0.9\linewidth]{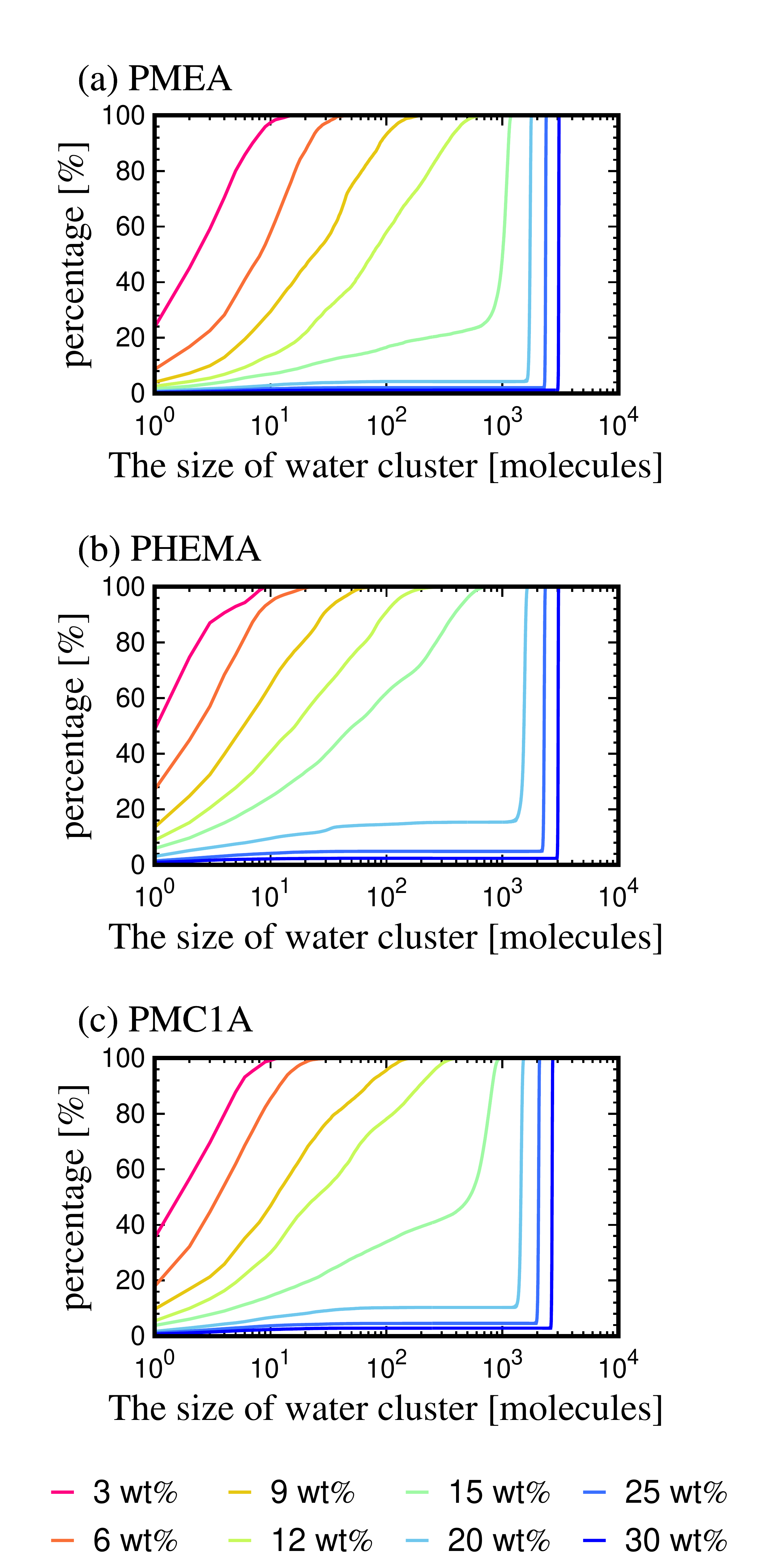}
\caption{Accumulated percentage of water molecules as a function of the
 size of the water cluster for (a) PMEA,
 (b) PHEMA, and (c) PMC1A, at various water contents up to 30 wt\%.
Note that as the water content increases, a
 step-wise behavior is observed at sizes of the water cluster larger
 than $10^3$ molecules,
 indicating the formation of a single cluster, similar to that 
 in bulk water.
}
\label{fig:watercluster}
\end{figure}

However, the characterization of the time scale of H-bond rearrangements
in individual water molecules has yet to be thoroughly scrutinized. 
This study aims to shed light on the dynamics of H-bonding of water
molecules confined within PMEA by focusing on the acceptor oxygen as the
bonding counterpart and categorizing the dynamic properties of water
molecules.
The impact of side-chain terminal groups and side-chain length on
H-bonding and water molecule diffusion is also examined through
simulations of hydrated PMEA analogues, such as poly(2-hydroxyethyl
methacrylate) (PHEMA) and poly(1-methoxymethyl acrylate) (PMC1A) by
varying the water content.
The diffusivity of water molecules confined within polymers is
further characterized and compared with that of bulk supercooled water.~\cite{gallo2016Water}
The study analyzes how the confinement effect enhances the cage-effect and
non-Gaussian behavior in single-particle displacement, which are frequently
observed in supercooled water.
Furthermore, the correlation between the diffusion of water molecules and
H-bond breakage is analyzed.
Therefore, this study presents a comprehensive discussion of the effects of the 
functional group on H-bonding and water molecule diffusion in hydrated
PMEA and its analogues.

\section{Simulation details}

The structure of the polymer studied in this paper is shown in
Fig.~\ref{fig:polymer}.
In this study, we conducted MD simulations of PMEA and its structural
analogues, PHEMA and PMC1A.
Each polymer features a side-chain possessing three oxygen atoms,
namely, the carbonyl oxygen ({\ob}), ether-like oxygen( {\os}), and
methoxy or hydroxy oxygen ({\om}), as denoted in Fig.~\ref{fig:polymer}.
The structure of PHEMA is different from that of PMEA in that it has a methyl
group on the backbone and a hydroxyl group at the terminal of the side
chain instead of a methoxy group. 
PMC1A, on the other hand, is different from PMEA in terms of the
number of methylene groups between the {\os} and {\om} sites.
It is worth mentioning that both PHEMA and PMC1A exhibit inferior blood
compatibility compared to PMEA and lack intermediate water
molecules.~\cite{tanaka2002Study, kobayashi2017Poly}

Each polymer was created by J-OCTA~\cite{JOCTA} using the OPLS-AA force
field.~\cite{jorgensen1996Development}
The polymer chain was atactic with a 1:1 steric control and a 
degree of polymerization of 50.
A hydrogen atom was placed at each terminal of the polymer molecules,
and number of polymer molecules in the simulation box was 20.
Water molecules were modeled using the TIP4P/2005 model~\cite{abascal2005General} and were added to
the box to achieve mass water contents ranging from 3 to 90
wt\% under periodic boundary conditions.
Note that the saturated water content of PMEA in the experimental
conditions has been 
reported to be 9 wt\%.~\cite{tanaka2000Cold}
MD simulations were performed using GROMACS~\cite{abraham2015GROMACS}
and initial molecular configurations were created using
PACKMOL.~\cite{martinez2009PACKMOL}
Initially, the $NPT$ ensemble calculation was performed for 10 ns at a
temperature of 300 K and pressure of 1 bar. 
For PHEMA, a subsequent $NPT$
ensemble simulation was conducted for 10 ns at 300 K and 1000 bar to
eliminate any cavities that may have arisen from the high hydrophilicity
of PHEMA during the first step $NPT$ calculation.
The $NVT$ ensemble was then utilized for a 5 ns annealing at 1000
K. 
Afterward, the system's box size was determined via $NPT$ equilibration
at 300 K and 1 bar, followed by a product run in the $NVT$ ensemble for up
to 500 ns at 300 K. 
The time step was set to 1 fs, and 
temperature and pressure were controlled using the Nos{\'e}--Hoover
thermostat~\cite{nose1984Unified, hoover1985Canonical} and the
Parrinello--Rahman barostat,~\cite{parrinello1981Polymorphic} respectively.

\section{Results and discussion}

\subsection{H-bond definition and size of water cluster}

\begin{figure*}[t]
\centering
\includegraphics[width=0.9\linewidth]{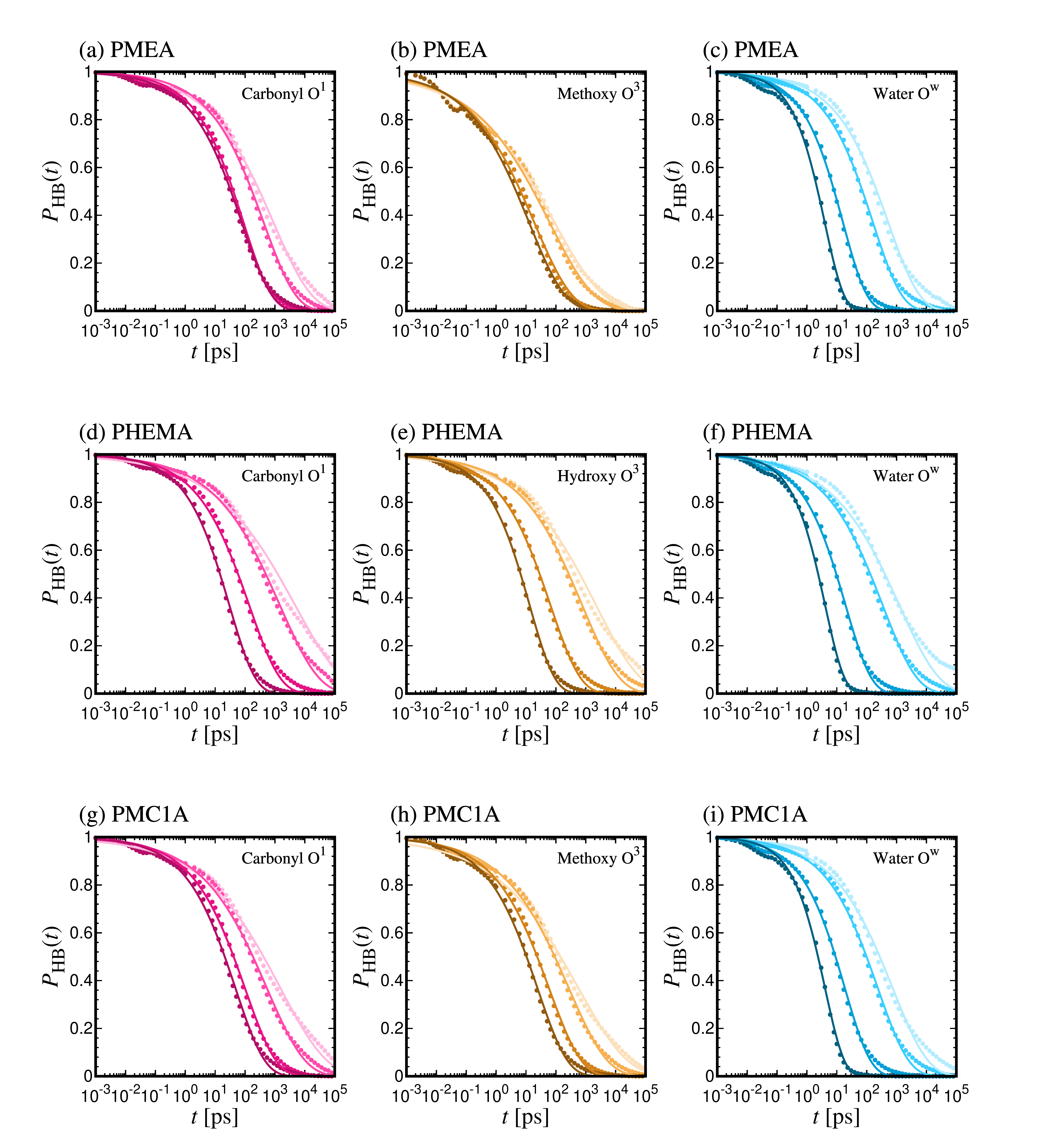}
\caption{H-bond time correlation function $\phb(t)$ with 
 respect to acceptor oxygen (carbonyl {\om}, methoxy or hydroxy {\ob}, and water {\ow}) 
[(a)-(c) PMEA, (d)-(f) PHEMA,  and (g)-(i) PMC1A].
The solid line represents the result of fitting with the stretched
 exponential function, $\phb(t) \simeq
\exp[-(t/\tau_{\mr{KWW}})^{\beta_{\mr{KWW}}}]$.
The results are shown for water content varying from right to left: 3
 wt\%, 9 wt\%, 30 wt\%, and 90 wt\% in each panel, with the color being darker at larger water contents.
}
\label{fig:phb}
\end{figure*}

\begin{figure*}[t]
\centering
\includegraphics[width=0.9\linewidth]{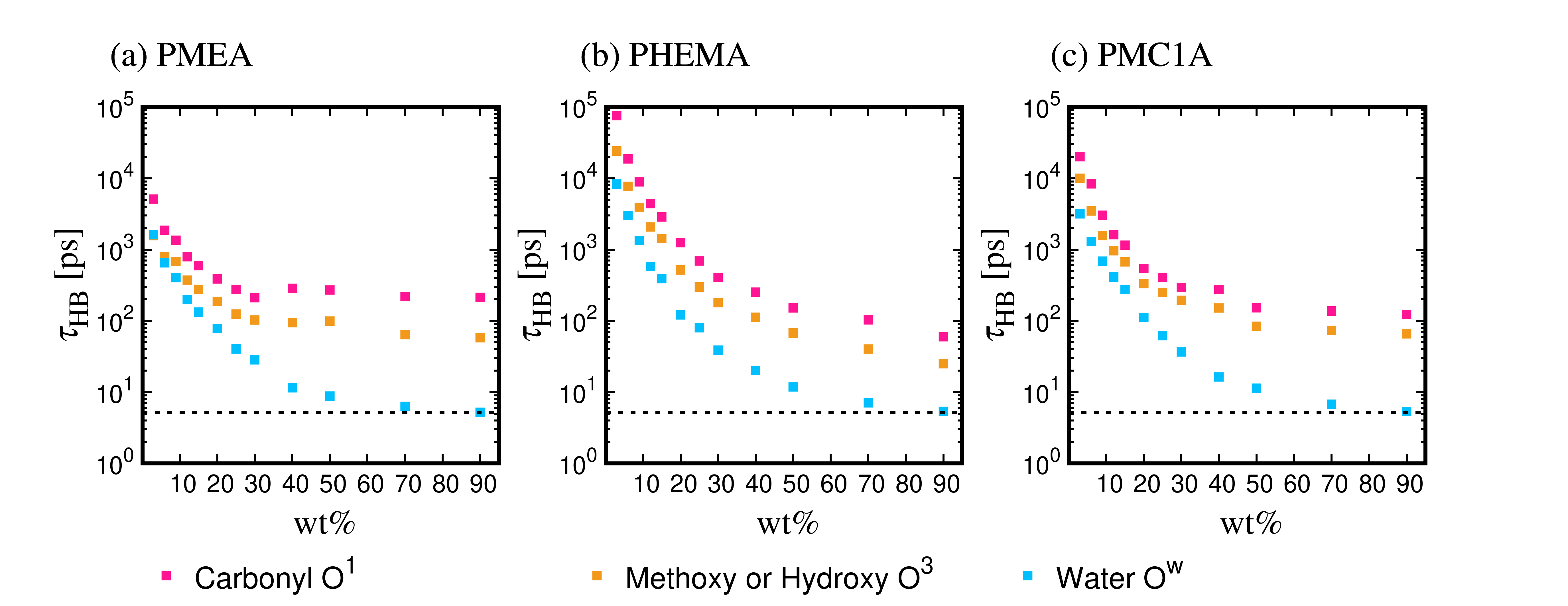}
\caption{
Water content dependence of 
H-bond lifetime $\thb$ for each acceptor oxygen ({\ob},  {\om}, and {\ow}) in
 (a) PMEA, (b) PHEMA, and (c) PMC1A.
The dashed line represents 
$\thb$ in bulk water at 1 g/cm$^3$ and 300 K. 
}
\label{fig:tau}
\end{figure*}

To examine the state of H-bond in water molecules confined
within polymers, we calculated the distance and angle distribution
function,
$g(r_\mr{oo},\beta)$, using the O-O distance, $r_\mr{oo}$, and the
angle of O-OH, $\beta$.~\cite{kumar2007Hydrogen, kikutsuji2018How, kikutsuji2019Consistency}
The expression $2\pi\rho r_\mr{oo}^2 \sin\beta g(r_\mr{oo}, \beta)\dd{r_\mr{oo}}\dd{\beta}$ 
represents the averaged
number of oxygen atoms present in the partial spherical shell with
$\dd{r_\mr{oo}}$ and $\dd{\beta}$, located at a distance $r_\mr{oo}$ and
angle $\beta$ from one of the hydrogen atoms of the donor, where
$\rho$ is the molecular number density of the system.
We then derived a 2-dimensional (2D) potential of mean force (PMF) between water molecule
oxygen ({\ow}) and each
polymer oxygen ({\ob}, {\os}, {\om}) or water molecule oxygen ({\ow}),
by setting $W(r_\mr{oo},\beta) = -k_\mr{B}T\ln{g(r_\mr{oo},\beta)}$.

Figure~\ref{fig:pmf_pmea} displays the 2D PMF $W(r_\mr{oo},\beta)$ of
PMEA-water system at 9 wt\%.
The results of the PHEMA and PMC1A systems are presented in Fig.~S1  and
Fig.~S2 of supplementary material, respectively.
The 2D PMF remains unaltered upon variation in water content for
each polymer system (data not shown).
As indicated in Fig.~\ref{fig:pmf_pmea}, the state of H-bond is
determined 
by the energetically stable region, which is defined as 
$(r_\mr{OO}, \beta) \le (0.35~\mathrm{nm}, 30^\circ)$.
It is noteworthy that
the most stable region between {\ow} and ether-like oxygen {\os} was
observed in the second coordination shell outside the H-bond region, and
the common tendency is seen with PHEMA and PMC1A in Figs.~S1(b) and
S2(b).
Henceforth, 
we consider the carbonyl oxygen ({\ob}), the methoxy or hydroxy
oxygen ({\om}), and the water oxygen ({\ow}) as H-bond 
acceptor oxygen.

To understand the water dynamics in the confined systems,
it is crucial to characterize the size of water clusters in the polymer-water system.
We calculated the size of water cluster using a criterion based on the H-bond
length, as used in a previous study.~\cite{kuo2019Analyses}
Water molecules are
considered to belong to the same cluster if they satisfy H-bond
length of $r_\mr{oo} \le 0.35$ nm.
Figure~\ref{fig:watercluster} shows the accumulated percentage of water
molecules as a function of the size of the water cluster at various
water contents.
All polymer-water systems exhibit small clusters of less than 10
water molecules at 3 wt\%.
As the water content increases, the water molecules begin to aggregate,
resulting in an increase in the cluster size.
Eventually, the clusters connect with each other to form large clusters of
approximately $10^3$ molecules at 30 wt\%.
Interestingly, PHEMA shows a smaller cluster size at water contents
below 15 wt\% compared to PMEA and PMC1A. 
This is attributed to the hydrophilic nature of PHEMA, which hinders water
molecules clustering.
The smaller cluster size results in
a lower porosity, which may contribute to lower mobility of water
molecules compared to PMEA
and PMC1A.

\subsection{H-bond lifetime}

The H-bond time correlation function is represented as
\begin{align}
\phb(t) = \dfrac{\left\langle h_{i,j}(t)h_{i,j}(0)\right\rangle}{\left\langle h_{i,j}(0)\right\rangle},
\label{eq:phb}
\end{align}
where $h_{i,j}(t)$ denotes the H-bond operator, such that
$h_{i,j}(t)$ equals unity if the water molecule $i$ acting as the donor
and the acceptor oxygen $j$ is H-bonded at 
time $t$ and zero if the bond is not present.~\cite{rapaport1983Hydrogen,
luzar1996Hydrogenbonda, luzar1996Effect, luzar2000Resolving}
The symbol $\langle \cdots \rangle$ represents the ensemble average over
all possible pairs of H-bonds at the initial time 0.
The $\phb(t)$ was calculated for 
each acceptor oxygen, and the results are shown in Fig.~\ref{fig:phb}.
The relaxation of $\phb(t)$ 
significantly slows down 
as the water content decreases and is found to exhibit a stretched exponential decay.
Thus, $\phb(t)$ was fitted by the Kohlrausch–Williams–Watts (KWW) function, 
$\phb(t) \simeq
\exp[-(t/\tau_{\mr{KWW}})^{\beta_{\mr{KWW}}}]$.
The exponent $\beta_\mr{KWW}(<1)$ characterizes how much the decay is
stretched compared to the exponential decay of $\beta_{\mr{KWW}} = 1$.
The values of the exponent $\beta_\mr{KWW}$ are displayed in Fig.~S3
of supplementary material.
For comparison, we also calculated 
the $\phb(t)$ of TIP4P/2005 liquid water system at 1 g/cm$^3$ and
temperatures ranging from 300 K to 200 K.
Figure~S4(a) and S4(b) of supplementary material display the temperature
dependence of $\phb(t)$ and $\beta_\mr{KWW}$, respectively.

The integral of $\phb(t)$ characterizes the H-bond lifetime $\thb$, 
\begin{align}
\tau\hb = \int  \phb(t) \dd{t}.
\label{eq:intkww}
\end{align}
By using the stretched exponential form, $\thb$ was estimated by the
mean relaxation time as 
$\tau\hb \approx ({\tau_{\mr{KWW}}}/{\beta_\mr{KWW}})\Gamma(1/\beta_\mr{KWW})$
with the Gamma function $\Gamma(\cdots)$.
It should be noted that when $\beta_\mathrm{KWW}$ equals to 1, 
$\tau\hb$ is the samel as $\tau_\mathrm{KWW}$.

The water content dependence of $\thb$ is displayed in \fig{tau}.
{Overall, as the water content decreases,
$\thb$ increases in all polymer-water systems, indicating a slowing down in
H-bond dynamics, which is similar to the time scale observed in 
supercooled water (see
Fig.~S4(c) of supplementary material that provides the temperature
dependence of $\thb$ in bulk water).
On the other hand, as the water content increases, 
$\thb$ ultimately reached a constant value.
Note that $\thb$ of {\ow} is nearly equal to that 
of bulk water at 1 g/cm$^3$ and 300 K, $\thb \approx 5$ ps.
Furthermore, the values of $\thb$ were found to be ranked in the order of {{\ow}$<$
{\om} $<$ {\ob}}.
The hydrophilicity of the functional groups notably enhances
$\thb$ at low water contents.
Specifically, in PHEMA, $\thb$ becomes approximately 10 times larger
than that of PMEA and PMC1A, indicating the strong affinity of
water to PHEMA resulting from the persistent 
H-bond between water molecules and the polymer.
In contrast, in the PMEA system, 
$\phb(t)$ for the methoxy
{\om} in PMEA displays a relaxation faster than 
the other acceptors, the carbonyl {\ob} and water 
{\ow}, particularly at water contents lower than 9 wt\%, as shown in Fig.~\ref{fig:phb}(a)-(c).
Note that the methoxy {\om} has a smaller value of
$\beta_\mr{KWW}$ compared to the other
acceptors (see Fig.~S3(a) of supplementary material).
A more stretched exponential decay of {\om} results in $\thb$ of {\om} that is comparable to that of {\ow}
at low water contents.
These observations demonstrate 
unique H-bonding interactions between water molecules and the methoxy
gourp of PMEA.

\subsection{Classification of water molecule states}

\begin{table}[b]
\caption{Classification of water molecules}
\label{tab:bw}
\begin{center}
\begin{tabular}{clll} 
\toprule
\multicolumn{2}{c}{\multirow{2}{*}{Classification}} &
 \multicolumn{2}{c}{acceptor oxygen species}  \\ \cline{3-4}
&&\multicolumn{1}{c}{1}&\multicolumn{1}{c}{2} \\ \midrule
UW &(Unstable Water) & N & \quad (N, $\mathrm{O}^3, \mathrm{O}^\mathrm{w}, \mathrm{O}^1$)\\
FDW &(Fast Dynamics Water) & $\mathrm{O}^3$& \quad($\mathrm{O}^3, \mathrm{O}^\mathrm{w}, \mathrm{O}^1$)\\
IDW &(Intermediate Dynamics Water) & $\mathrm{O}^\mathrm{w}$ & \quad ($\mathrm{O}^\mathrm{w}, \mathrm{O}^1 $)\\
SDW &(Slow Dynamics Water) & $\mathrm{O}^1$& \quad $\mathrm{O}^1$ \\ \bottomrule
\end{tabular}
\end{center}
\end{table}

\begin{figure*}[t]
\centering
\includegraphics[width=0.9\textwidth]{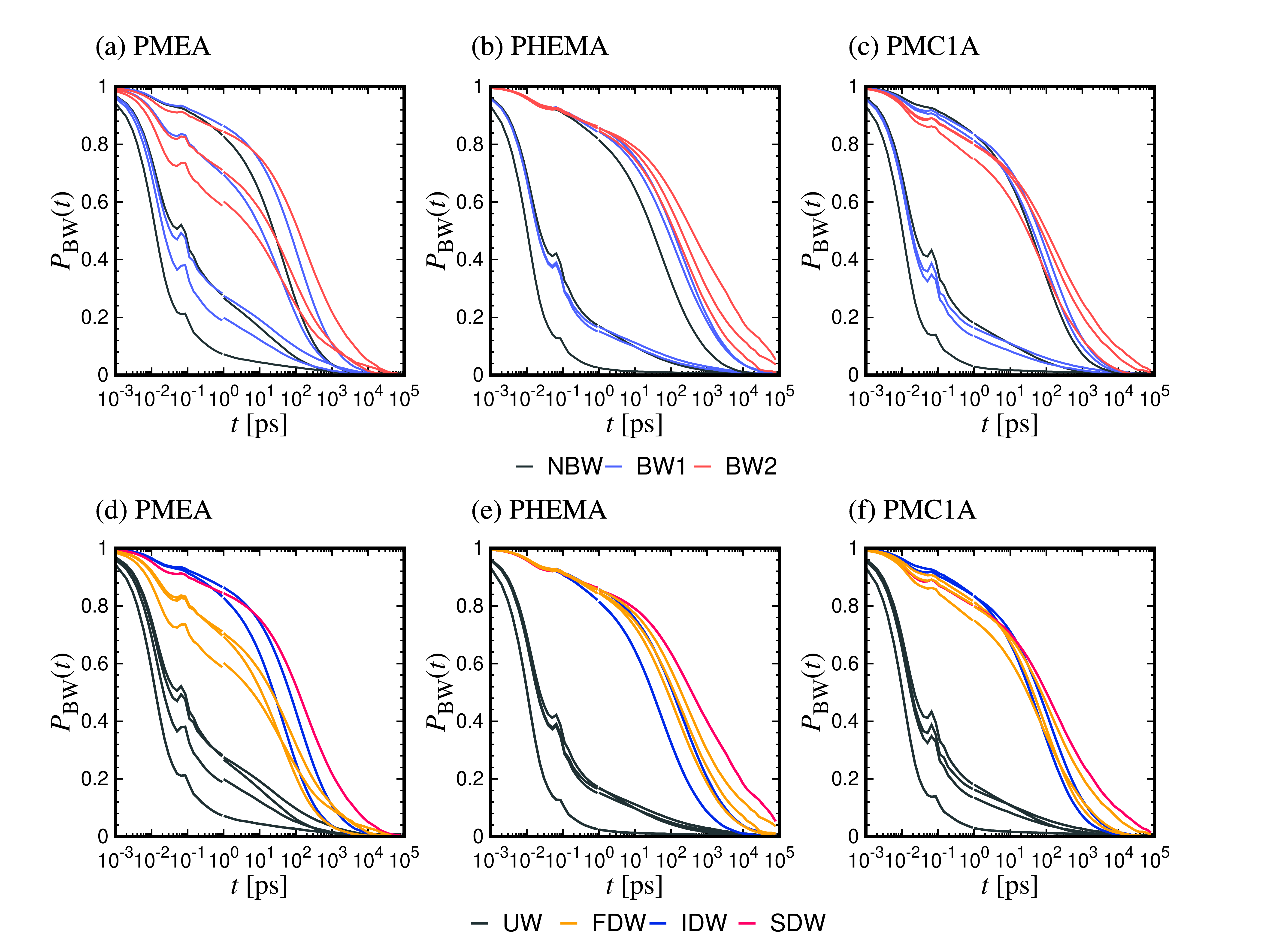}
\caption{Relaxation of 10 water molecule states characterized by the
 time correlation function $\pbw(t)$ at 9 wt\% in PMEA [(a) and (d)], PHEMA [(b)
 and (e)], and PMC1A [(c) and (f)].
(a)-(c) Classification by the number of H-bons with polymeric oxygen,
 namely, NBW, BW1, and BW2 (see Refs.~\onlinecite{kuo2020Elucidating,
 kuo2020Molecular, kuo2021Effects}), are represented by black, blue, and
 red
 color, respectively.
Since NBW, BW1, and BW2 represent the number of H-bonds with the
 polymer, they correspond to 3, 4, and 3 curves, respectively. 
(d)-(f) Classification by the acceptor oxygen of water molecule, namely, UW,
FDW,  IDW, and SDW (see \tab{bw}), are represented by black, orange,
 blue, and red
 color, respectively.
In total, the 10 curves are the same between (a) and (d), (b) and (e),
 and (c) and (f). The difference is the classification scheme
 represented by color codes.
}
\label{fig:bw}
\end{figure*}

\begin{figure*}[t]
\centering
\includegraphics[width=0.9\textwidth]{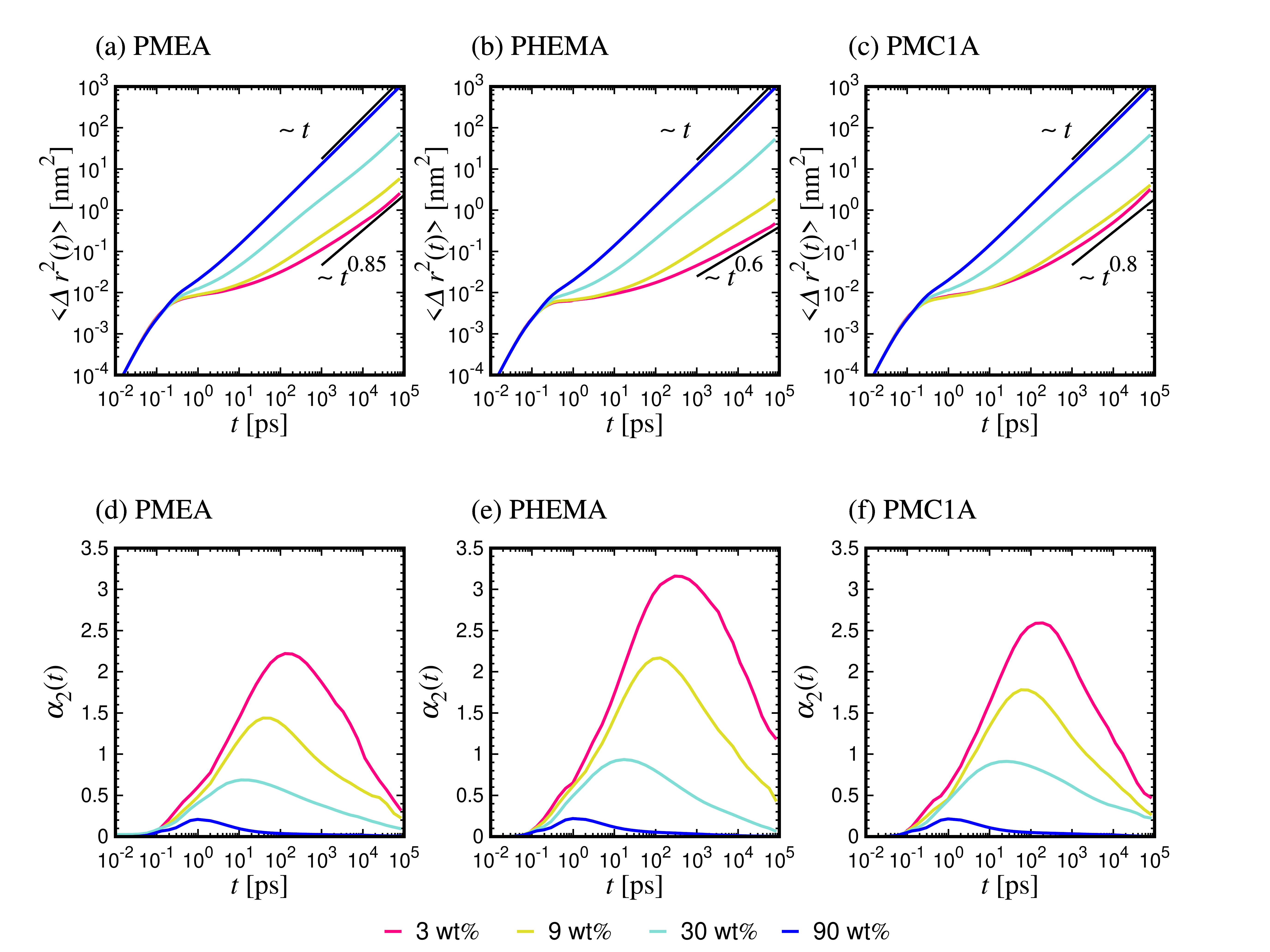}
\caption{(a)-(c) MSD $\langle \Delta r^2(t) \rangle$ of water molecules in (a) PMEA, (b) PHEMA, and (c) PMC1A.
For viewing guide, the diffusive and sub-diffusive behaviors, $\langle \Delta r^2(t)
 \rangle \sim t^\alpha$, are represented by black lines with $\alpha=1$ at 90 wt\%,
and $\alpha=$0.85 (PMEA), 0.6 (PHEMA) and 0.8 (PMC1A) at 3 wt\%, respectively.
(d)-(f) NGP $\alpha_2(t)$ of water molecules in (d) PMEA, (e) PHEMA, and (f) PMC1A.
}
\label{fig:msd}
\end{figure*}

\begin{figure}[t]
\centering
\includegraphics[width=0.9\linewidth]{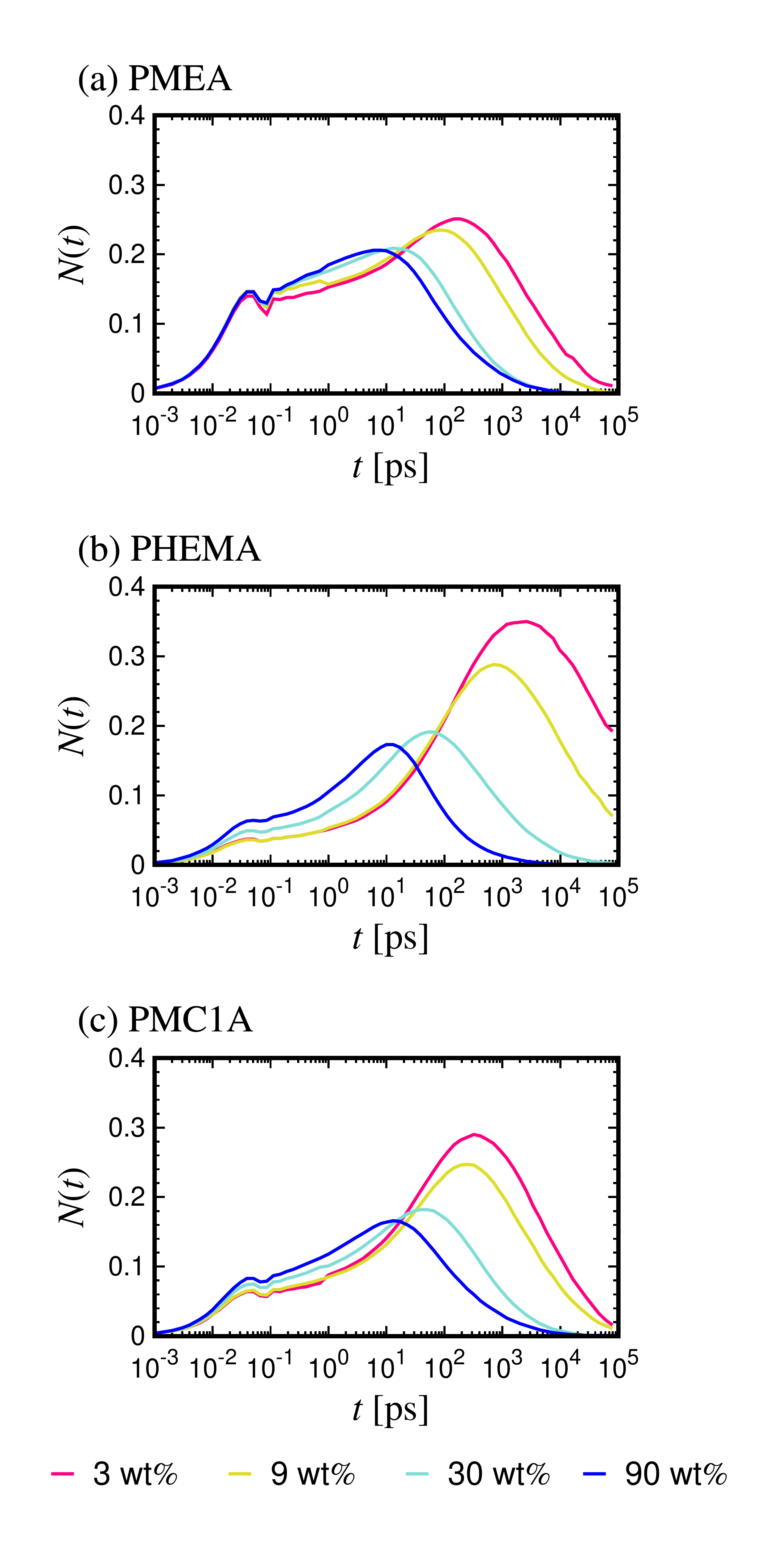}
\caption{Time correlation function $N(t)$ evaluating 
the contribution of the water molecule that does not diffuse after the
 breakage of H-bond with {\om} in (a) PMEA, (b) PHEMA, and (c) PMC1A.
}
\label{fig:pdr}
\end{figure}        

The H-bond time correlation function $\phb(t)$ is further examined to elucidate
the dynamics of water molecules in more detail.
Each water molecule has four possible states based on its H-bond
acceptor oxygen (or the absence), namely {\ob}, {\om}, and {\ow}, or no H-bond (N).
Since a single water molecule has two donor hydrogen atoms,
there are a total of 10 possible states for each water molecule
when considering the overlapping combinations.
Thus, the H-bond operator is expanded to 
$h_{i,j,k}(t)$, which is unity when water molecule $i$ is H-bonded with 
acceptor oxygen $j$ and also H-bonded with acceptor oxygen $k$ at 
time $t$ and zero otherwise.
Note that 10 states can be classified into three water molecules based on
the number of H-bonds with the polymer: NBW, BW1, and
BW2 (note that the H-bond with {\ow} is not counted in the classification
into NBW, BW1, and BW2)~\cite{kuo2020Elucidating, kuo2020Molecular,
kuo2021Effects}.
The time correlation function of $h_{i,j,k}(t)$ is analogous to
\eq{phb} and expressed by 
\begin{align}
\pbw(t) = \dfrac{\left\langle h_{i,j,k}(t)h_{i,j,k}(0)\right\rangle}{\left\langle h_{i,j,k}(0)\right\rangle}.
\label{eq:phb10}
\end{align}
At 9 wt\%, 
the results of $\pbw(t)$ of 10 water molecule states are plotted in
\fig{bw}.
In addition, Figs.~S5, S6, and S7 of supplementary materials show
$\pbw(t)$ at other water contents, 3 wt\%, 30 wt\%, and 90 wt\%, respectively.

The results show that 
the relaxation curves can be categorized into two modes: a slow
relaxation mode and a fast relaxation mode with a small tail.
Additionally, the intermediate relaxation mode was observed in PMEA's $\pbw(t)$.
However, 
the NBW, BW1, and BW2 classification (as shown in
\fig{bw}(a)-(c))
 is insufficient to distinguish between the relaxation modes.
The relaxation of BW1 may overlap with those of BW2 and NBW,
and BW2 and NBW correspond roughly to
fast and slow modes, respectively.
Moreover, the intermediate relaxation mode observed in
PMEA's $\pbw(t)$ requires further clarification.
A more detailed classification may be
necessary, and one possible approach is to compare the water molecule
states in PMEA with those in PHEMA and PMC1A.

We propose a novel classification for $\pbw(t)$ based on the H-bond
lifetime time $\thb$, taking into account the significant difference in 
$\thb$ between {\ob} and {\om} in PMEA (see \fig{tau}(a)).
The new classification, presented in \tab{bw}, 
mainly classifies acceptor oxygen species 1 in the order of N, {\om}, {\ow}, and
{\ob}.
It results in three stable types of water: FDW (Fast Dynamics
Water), IDW (Intermediate Dynamics Water), and SDW (Slow Dynamics
Water) based on H-bonds with {\om}, {\ow}, and
{\ob}, respectively.
We also define an 
Unstable Water (UW) state where at least
one of the donor hydrogen atoms does not form an H-bond, causing the
state to relax very rapidly.

As shown in \fig{bw}(d)-(f), 
our proposed classification better characterizes 
the relaxations behavior of $\pbw(t)$.
The SDW state displays the slowest relaxation due to its two H-bonds with {\ob}.
While the relaxation of IDW is comparable to that of SDW, 
IDW shows a greater dependence on the water content when compared 
to SDW (see Figs.~S5-S7 of supplementary material).
Moreover, our results indicate that 
the relaxation of FDW, which is attributed to the H-bond with {\om}, is
faster in PMEA compared to SDW and IDW.
It is important to note that the H-bond lifetime $\thb$ of PMEA was
significantly different between {\om} and {\ob} (see again \fig{tau}(a)), which should be
considered to classify $\pbw(t)$.
Besides, FDW displays a relaxation behavior more analogous to that of SDW and IDW in
the cases of PHEMA and PMC1A.

\subsection{Water molecule diffusion}

The mean square displacement (MSD) was calculated to evaluate the
diffusion of water molecules in each polymer-water system.
The MSD is expressed by 
\begin{align}
\langle \Delta r^2(t) \rangle = \left\langle \dfrac{1}{N} \sum^N_{i=1}\left|\bm{r}_i(t) - \bm{r}_i(0) \right|^2\right\rangle,
\end{align}
where $\bm{r}_i(t) - \bm{r}_i(0)$ represents the displacement vector of
oxygen atom of water molecule $i$
between two times $0$ and $t$, and $N$ is the number of water molecules.

Figure \ref{fig:msd} presents the MSD of water molecules in each system
studied.
To provide a point of reference, 
Fig.~S8(a) of supplementary material displays the temperature dependence
of MSD in bulk supercooled water.
In the short time regime ($t < 10^{-1}$ ps), the MSD exhibits a 
proportional relationship with $t^2$, which indicate the ballistic
motion without colliding with other molecules.
The second regime is a time region of approximately $10^{-1}$ to $10^2$ ps,
where the MSD shows a plateau that is more prominent with decreasing water content.
This plateau represents the ``cage-effect'' observed also in supercooled water, where 
water molecules are confined by H-bonds for a significant duration (see
also Fig.~S8(a) of supplementary material).~\cite{gallo1996Slow,
sciortino1996Supercooled, giovambattista2004Dynamic, 
kawasaki2017Identifying, kawasaki2019Spurious,
kawasaki2019Classification, kikutsuji2019Diffusion, teboul2019Specific}
Finally, in the long-time regime, the cage-effect weakens and the
diffusivity of water molecules will be described by $\langle \Delta r^2(t) \rangle $
proportional to $t$.
However, the sub-diffusive behavior $\langle \Delta r^2(t) \rangle
\sim t^\alpha$ with $\alpha<1$ was observed at longer times.
This sub-diffusivity can be attributed to 
the low mobility of polymer side chains that create a heterogeneous environment, causing less water
molecule diffusion compared to that of bulk
water.~\cite{saxton1994Anomalous, bagchi2005Water, sung2008Lateral,
bellissent-funel2016Water, chong2016Anomalous, tan2018Graduala}
In fact, 
Fig.~\ref{fig:watercluster} shows that at water contents below
15 wt\%, most of water molecules are isolated and confined within
the polymer matrix.
This confinement effect is particularly pronounced in PHEMA due to 
its high hydrophilicity, resulting in the sub-diffusive behavior 
with the exponent $\alpha \approx 0.6$ at 3 wt\%, which is lower than
those observed in PMEA and PMC1A.
In contrast, the increase in the water content leads to the formation of water molecule clusters
and greater connectivity, as depicted in Fig.~\ref{fig:watercluster}.
Therefore, it is thought that the larger the connection of water
molecules, the easier it is to establish a path for diffusion.

The non-Gaussian parameter (NGP) $\alpha_2(t)$ was also calculated to assess
the deviation of the displacement of water molecules in polymers-water
system from the Gauss distribution.
The NGP is expressed by 
\begin{align}
\alpha_2(t) &= \dfrac{3}{5}\dfrac{\langle\D r^4(t)\rangle}{\left\langle\D \bm{r}^2(t)\right\rangle^2} -1,
\label{eq:ngp}
\end{align}
where
$\langle \D r^4(t) \rangle = \left\langle
(1/N)\sum_{i=1}^N\left|\bm{r}_i(t) - \bm{r}_i(0)\right|^4 \right\rangle$
is the fourth-order moment of displacement between 0 and $t$.
The NGP in supercooled water has been extensively studied and has been
found to exhibit a significant level of 
non-Gaussianity as the temperature decreases, suggesting the presence of
dynamic heterogeneity.~\cite{sciortino1996Supercooled, giovambattista2004Dynamic,
kawasaki2017Identifying, kawasaki2019Spurious, kawasaki2019Classification, teboul2019Specific}

The results of $\alpha_2(t)$ are presented in \fig{msd}.
The $\alpha_2(t)$ of bulk supercooled water is found in
Fig.~S8(b) of supplementary material.
The value of $\alpha_2(t)$ starts from 0 at short times, corresponding to the
ballistic regime in the MSD, and 
increases at intermediate times, where the MSD exhibits the plateau due
to the cage-effect.
The peak of $\alpha_2(t)$ is reached when escaping the cage-effect in the MSD.
The peak height of $\alpha_2(t)$ increases as the water content is lowered.
This indicates the significant non-Gaussianity of the displacement of water
molecules when water molecules are surrounded in the highly
heterogeneous environment created by the polymer matrix.
After the peak, $\alpha_2(t)$ converges to 0, corresponding to the diffusive behavior.
Interestingly, 
the order of the maximum peak height in $\alpha_2(t)$ and the
sub-diffusively of MSD corresponds to the
trend of the H-bond lifetime $\thb$, where PMEA has the shortest
lifetime and PHEMA has the longest. 
This suggests that the anomalous water dynamics confined within the
polymer matrix is related to the strength and lifetime of the H-bonds
between water and polymer.

\subsection{Correlation between H-bond breakage and water diffusion}

The correlation between diffusion of water molecules was investigated to
determine the impact of 
H-bond breakage on the diffusion of two H-bonded
water molecules.
In fact, H-bond networks can hinder the water diffusion.
Conversely, when two water molecules break an H-bond, 
they may form it again if they have not diffused away from each other.
To evaluate the contribution of 
the water molecule that does not diffuse after H-bond breakage, 
a time correlation function $N(t)$ was proposed, which is defined as 
\begin{align}
N(t) = \dfrac{\langle h_{i,j}(0)(1-h_{i,j}(t))H_{i,j}(t)\rangle}{\langle h_{i,j}(0)\rangle}.
\label{eq:nt}
\end{align}
Here, $H_{i,j}(t)$ is an index function that takes the value of unity if the distance
between oxygen atoms $i$ and $j$ is less than 0.35 nm and 0
otherwise.~\cite{luzar1996Hydrogenbonda, luzar1996Effect,
xu2001HydrogenBond, xu2002Cana, mondal2022Dynamic}
It should be noted that 
even if the O-O distance is close and $H_{i,j}(t)$ equals 1, 
$(1-h_{i,j}(t))$ is 1 when the H-bond is broken at $t$.
Consequently, $N(t)$ represents the conditional probability that a water
molecule and an acceptor oxygen remain in close proximity at time $t$,
even after H-bond breakage, provided that they were H-bonded at $t=0$.
At short times, $N(t)$ is anticipated to exhibit an ascending trend following the 
H-bond breakages, whereas 
$N(t)$ should approach to 0 due to water molecule diffusion at long times.

Figure~\ref{fig:pdr} depicts $N(t)$ of the water molecule that is
H-bonding with the methoxy or hydroxy oxygen ({\om}).
Results of $N(t)$ corresponding to the carbonyl oxygen ({\ob}) and water oxygen
({\ow}) can be found in Fig.~S9 of supplementary material.
Additionally, Fig.~S10 of supplementary material displays the temperature
dependence of $N(t)$ in bulk water for comparison.
As illustrated in \fig{pdr}, the peak of $N(t)$ occurs at
approximately the H-bond lifetime $\thb$, and 
becomes more prominent as water content decreased.
The maximum values of $N(t)$ at low water contents are higher for PHEMA,
followed by PMC1A and PMEA, corresponding to larger H-bond
lifetimes $\thb$ of {\om}.
This observation can be attributed to the low diffusivity of the water
molecules near polymers due to water and polymer interactions, despite the H-bond being broken.
In contrast, in supercooled water, as the temperature decreases, the
peak of $N(t)$ becomes less pronounced, 
indicating the relatively higher mobility of water molecules upon
breaking their H-bonds, rather than remaining in their original
positions, as demonstrated in Fig.~S10 of the supplementary material.
Notably, the behavior of $N(t)$ of the methoxy oxygen {\om} in PMEA exhibits distinct 
characteristics. 
As illustrated in \fig{pdr}(a), 
$N(t)$ exhibits a discernible 
fraction of $N(t) \approx 0.2$ persisting over a timescale of
approximately $1-10$ ps, irrespective of water content.
This observation suggests that following the H-bond breakages,
water molecules H-bonded with {\om} do 
not diffuse instantly, but rather exhibit rotational motions, indicating
weak binding of water molecules in the visinity of {\om} in
PMEA.
In contrast, the persistence of $N(t)$ on the picosecond timescale is not
observed in \fig{pdr}(b) and (c) for PHEMA and PMC1A, indicating that a water molecule
leaves the polymer surface upon breakage of the H-bond.

\section{Conclusion}

In this study, MD simulations were conducted to
investigate the dynamic properties of water molecules confined within
poly(meth)acrylates, including PMEA, PHEMA, and PMC1A. 
The mobility of water molecules confined in polymers was found to be significantly
slower than that in ordinary bulk water, and highly dependent on the
water content. 
As the water content decreases, the H-bond lifetime $\thb$
increases, 
approaching values observed in bulk supercooled water.
Additionally, 
the conventional classification of H-bond correlation function
based on the number of H-bonds with polymeric oxygen was
insufficient in describing the behavior of water molecules in PMEA. 
Instead, classifying the behavior of water molecules based on 
H-bond acceptors proved to be more effective.

The diffusive behavior of water molecules was analyzed by the MSD and
NGP methods, commonly employed in the study of supercooled water.
The results indicate 
the sub-diffusive and non-Gaussian behavior in
single particle displacements, which are attributed to 
the limited mobility of water molecules confined within the polymer matrix.}
Moreover, the H-bonds of the methoxy oxygen in
PMEA were observed to break more rapidly than 
in PHEMA and PMC1A at low to medium water contents, but
the water molecule and the methoxy group involved
remain in close proximity to one another. 


The characteristic
behaviors of water molecules in the proximity of the polymer
surface were observed through simulations and are
consistent with experimental results, which suggest 
the specific role of the PMEA methoxy group noted often within the
context of ``intermediate water''.~\cite{kitano2001Structure,
kitano2005Correlation, morita2007TimeResolved} 
Specifically, the present study has 
revealed that the H-bond dynamics of the methoxy
oxygen {\om} in PMEA, characterized by a time scale of $10^2$ - $10^3$ ps, is
one order of magnitude faster than that observed in other polymers, such
as PHEMA and PMC1A. 
Remarkably, the time scale of water molecules 
H-bonded with the methoxy oxygen {\om} in PMEA
approximates to that of
``intermediate water'' characterized by NMR spectroscopy, which may be
linked to a potential mechanism underlying protein denaturing adsorption.
Nevertheless, the precise relationship between the dynamics of
water molecules with intermediate time scales and blood compatibility of
PMEA remains to be elucidated.
Further studies are needed to fully understand the underlying mechanism
of protein denaturing adsorption, which is largely attributed to the
presence of water molecules in the proximity of the polymer surface.

\section*{Supplementary Material}
See supplementary material for 2D PMF of PHEMA and PMC1A
 (Figs.~S1 and S2), 
water content dependence of exponent $\beta_\mr{KWW}$ (Fig.~S3),
H-bond correlation function $\phb(t)$ and lifetime $\thb$ of
 bulk supercooled water (Fig.~S4),
 $\pbw(t)$ at water contents, 3 wt\%, 30 wt\%, and 90 wt\%
 (Figs.~S5, S6, and S7), 
MSD and NGP of bulk supercooled water (Fig.~S8)
$N(t)$ of carbonyl oxygen and
 water oxygen (Fig.~S9), and $N(t)$ of bulk supercooled water (Fig.~S10).

\begin{acknowledgments}
This work was supported by 
JSPS KAKENHI Grant-in-Aid for Scientific Research on Innovative Areas:
 Aquatic Functional Materials (No.~\mbox{JP22H04542}).
This work was also partially supported 
by the Fugaku Supercomputing Project (No.~\mbox{JPMXP1020200308}) and 
the Data-Driven Material Research Project (No.~\mbox{JPMXP1122714694})
from the
Ministry of Education, Culture, Sports, Science, and Technology.
K.K. acknowledges Professor Takashi Kato (University of Tokyo) and
 Professor Masaru Tanaka (Kyushu University) for valuable discussions.
The numerical calculations were performed at Research Center of
Computational Science, Okazaki Research Facilities, National Institutes
of Natural Sciences (Project: 21-IMS-C058).
\end{acknowledgments}

%

\section*{AUTHOR DECLARATIONS}

\section*{Conflict of Interest}
The authors have no conflicts to disclose.

\section*{Data availability statement}
The data that support the findings of this study are available from the corresponding author upon reasonable request.

%

\clearpage
\widetext

\setcounter{equation}{0}
\setcounter{figure}{0}
\setcounter{table}{0}
\setcounter{page}{1}

\renewcommand{\theequation}{S.\arabic{equation}}
\renewcommand{\thefigure}{S\arabic{figure}}
\renewcommand{\thetable}{S\arabic{table}}
\renewcommand{\bibnumfmt}[1]{[S#1]}
\renewcommand{\citenumfont}[1]{S#1}

\noindent{\bf\Large Supplementary Material}
\vspace{5mm}
\begin{center}
\textbf{\large  Revealing the hidden dynamics of confined water in acrylate
polymers:\\ Insights from hydrogen-bond lifetime analysis}
\\

\vspace{5mm}
 
{Kokoro Shikata,$^1$ Takuma Kikutsuji,$^1$ Nobuhiro Yasoshima,$^{1, 2}$ Kang Kim,$^1$
and Nobuyuki Matubayasi$^1$}
\\

\vspace{5mm}

\noindent
\textit{
$^{1)}$Division of Chemical Engineering, Graduate School of Engineering Science, Osaka University, Osaka 560-8531, Japan}
\textit{
$^{2)}$Department of Information and Computer Engineering, National
Institute of Technology, Toyota College, 2-1 Eiseicho, Toyota, Aichi,
471-8525, Japan}

\end{center}

\begin{figure}[htbp]
\centering
\includegraphics[width=0.75\linewidth]{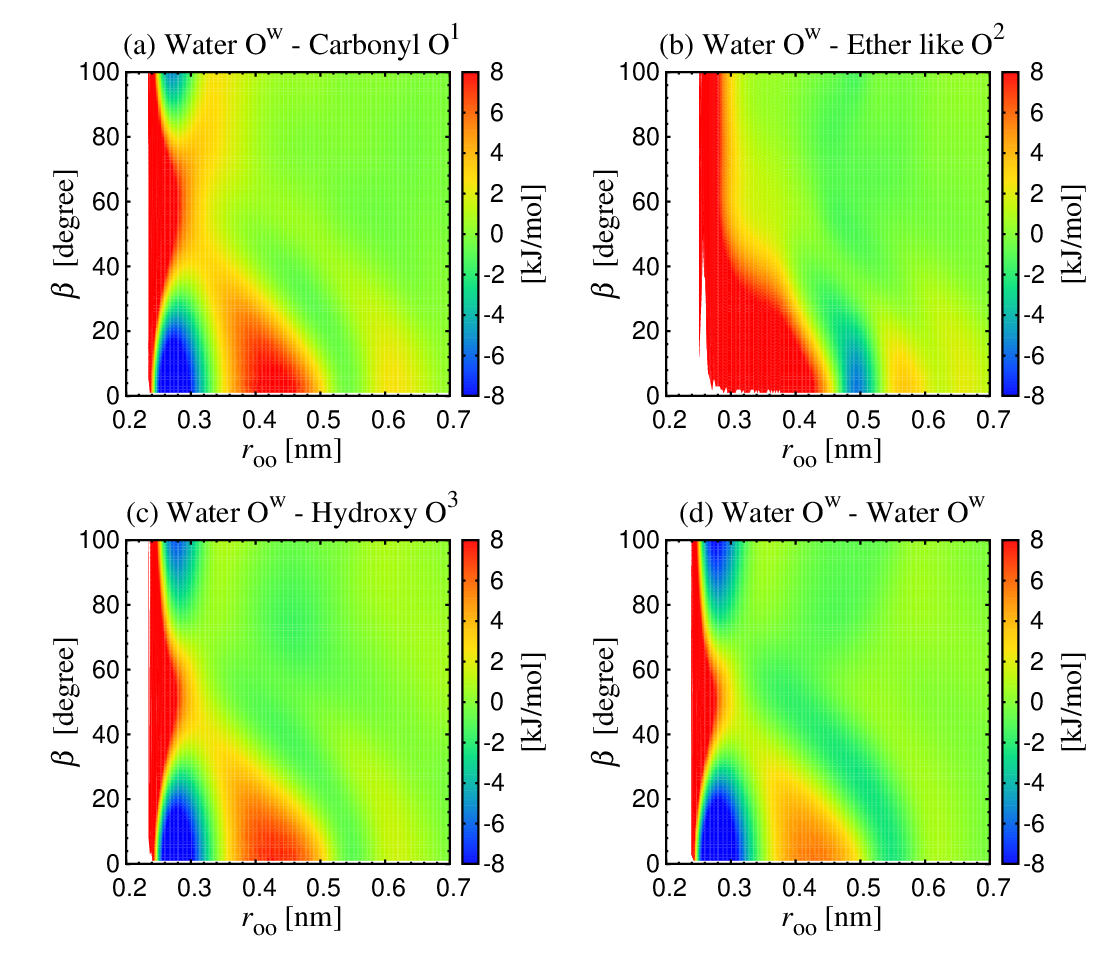}
\caption{2D PMF $W(r_\mr{oo},\beta)$ between water oxygen ({\ow})
 and acceptor oxygen [(a) carbonyl {\ob}, (b) ether-like {\os}, (c) hydroxy {\om},
 and (d) water {\ow}] in PHEMA-water system
 at 9 wt\%.}
\end{figure}

\begin{figure}[htbp]
\centering
\includegraphics[width=0.75\linewidth]{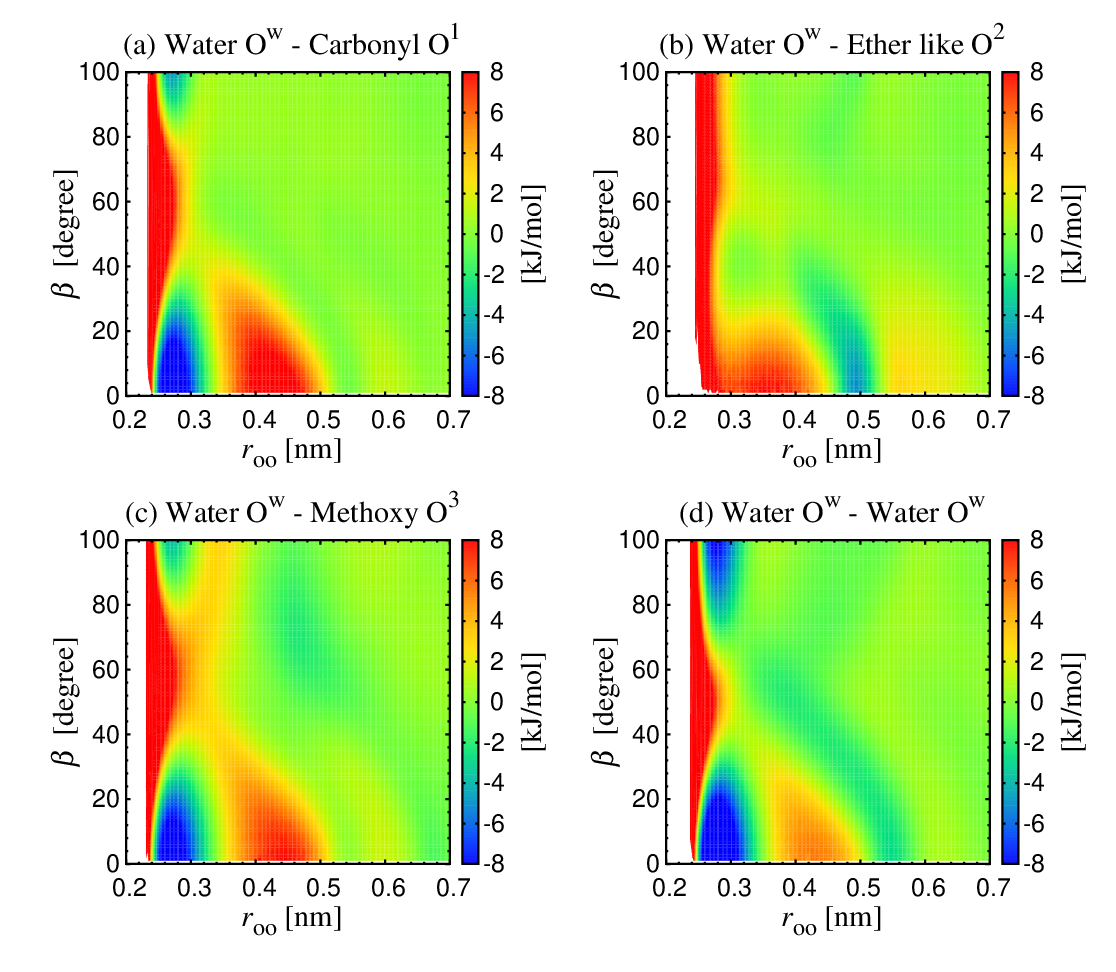}
\caption{2D PMF $W(r_\mr{oo},\beta)$ between water oxygen ({\ow})
 and acceptor oxygen [(a) carbonyl {\ob}, (b) ether-like {\os}, (c) methoxy {\om},
 and (d) water {\ow}] in PMC1A-water system
 at 9 wt\%.}
\end{figure}

\begin{figure}[htbp]
\centering
\includegraphics[width=0.9\linewidth]{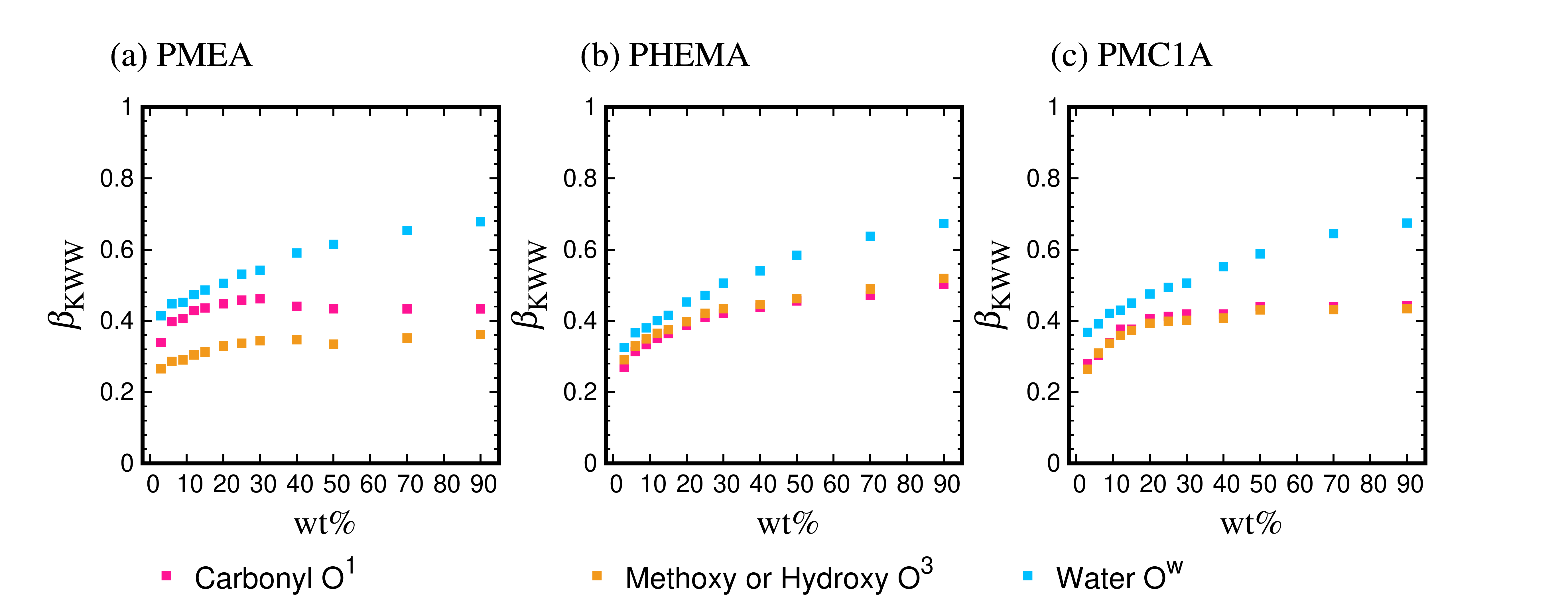}
\caption{Water content dependence of $\beta_\mathrm{KWW}$ for each
 acceptor oxygen (carbonyl \ob, methoxy or hydroxy \om, and water \ow) in (a) PMEA, (b) PHEMA, and (c) PMC1A.}
\end{figure}

\begin{figure}[htbp]
     \centering
     \includegraphics[width=0.9\linewidth]{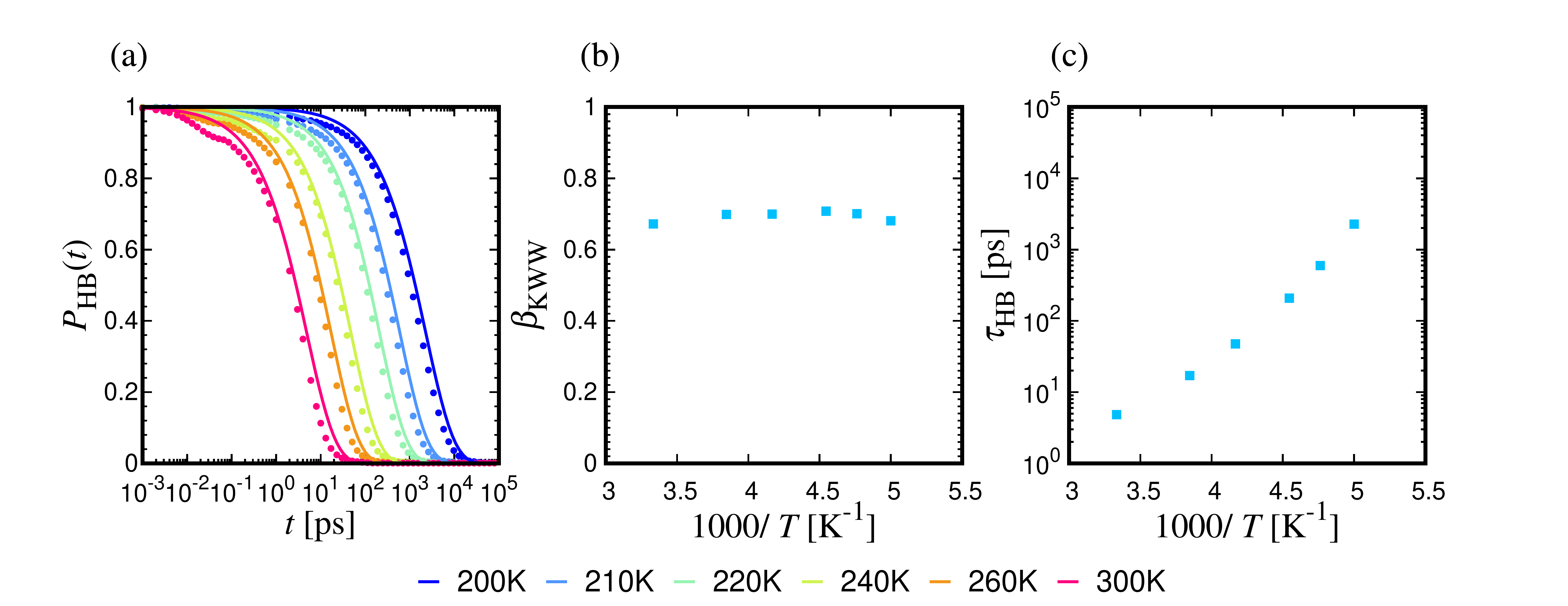}
     \caption{(a) H-bond time correlation function $\phb(t)$ in
 TIP4P/2005 liquid water system at 1 g/cm$^3$.
The solid line represents the result of fitting with the stretched
 exponential function, $\phb(t) \simeq
\exp[-(t/\tau_{\mr{KWW}})^{\beta_{\mr{KWW}}}]$.
Temperature dependence of $\beta_\mathrm{KWW}$ (b) and $\thb$ (c) .}
\end{figure}

\begin{figure}[htbp]
\centering
\includegraphics[width=0.9\linewidth]{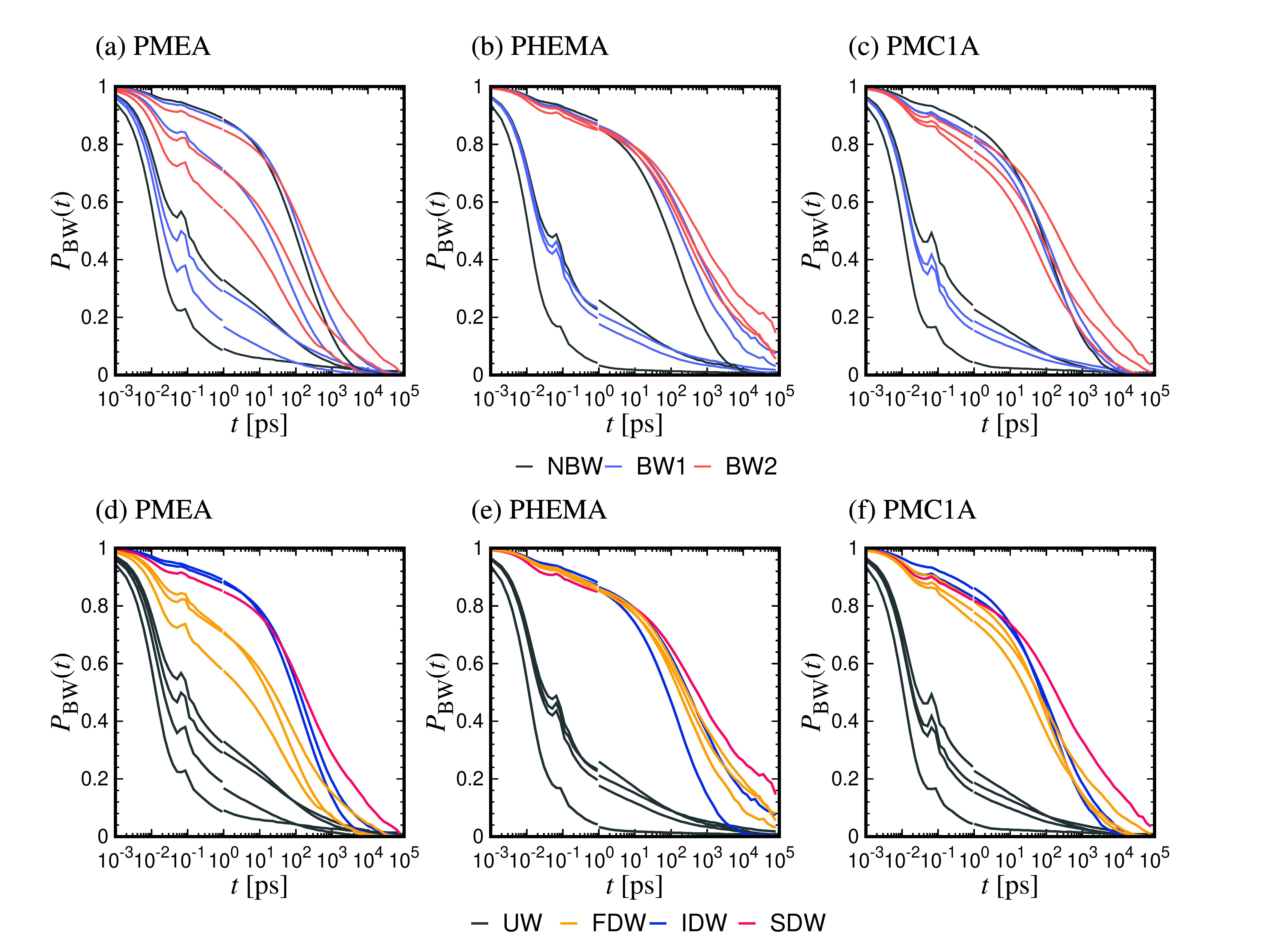}
\caption{Relaxation of 10 water molecule states characterized by the
 time correlation function $\pbw(t)$ at 3 wt\% in PMEA [(a) and (d)],
 PHEMA [(b) and (e)], and PMC1A [(c) and (f)]. 
(a)-(c) Classification by the number of H-bonds with polymeric oxygen,
     namely, NBW, BW1, and BW2 are represented by black, blue, and red
     color, respectively.
Since NBW, BW1, and BW2 represent the number of H-bonds with the
 polymer, they correspond to 3, 4, and 3 curves, respectively. 
(d)-(f) Classification by acceptor oxygen of water molecule, namely, UW,
     FDW, IDW, and SDW, are represented by black, orange, blue, and red
     color, respectively.
In total, the 10 curves are the same between (a) and (d), (b) and (e),
 and (c) and (f). 
The difference is the classification scheme
 represented by color codes.
}
\end{figure}

\begin{figure}[htbp]
\centering
\includegraphics[width=0.9\linewidth]{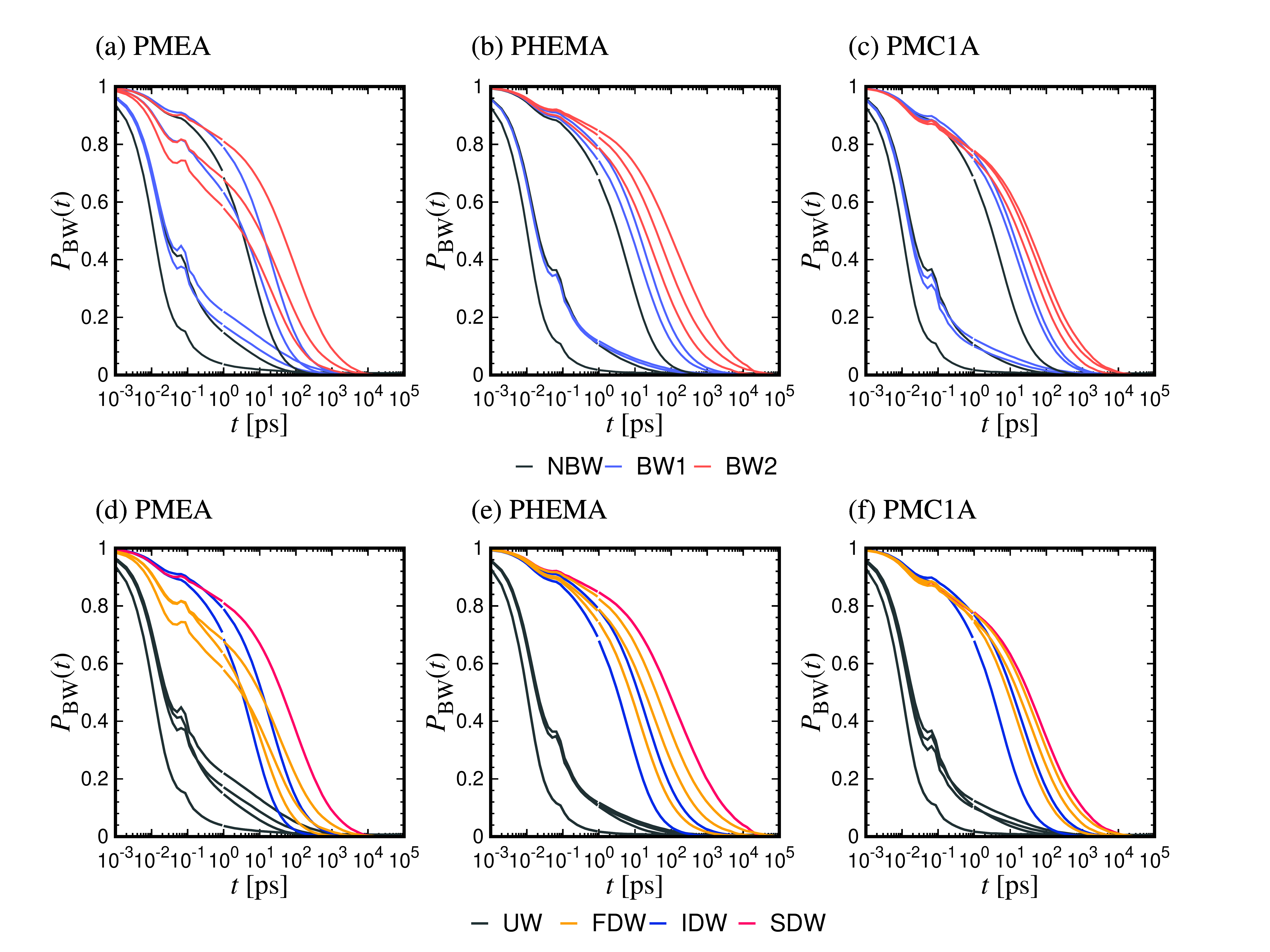}
\caption{Relaxation of 10 water molecule states characterized by the
 time correlation function $\pbw(t)$ at 30 wt\% in PMEA [(a) and (d)],
 PHEMA [(b) and (e)], and PMC1A [(c) and (f)].
(a)-(c) Classification by the number of H-bonds with polymeric oxygen,
 namely, NBW, BW1, and BW2 are represented by black, blue, and red
 color, respectively.
Since NBW, BW1, and BW2 represent the number of H-bonds with the
 polymer, they correspond to 3, 4, and 3 curves, respectively. 
(d)-(f) Classification by acceptor oxygen of water molecule, namely, UW,
 FDW, IDW, and SDW, are represented by black, orange, blue, and red
 color, respectively.
In total, the 10 curves are the same between (a) and (d), (b) and (e),
 and (c) and (f). 
The difference is the classification scheme
 represented by color codes.
}
\end{figure}

\begin{figure}[htbp]
\centering
\includegraphics[width=0.9\linewidth]{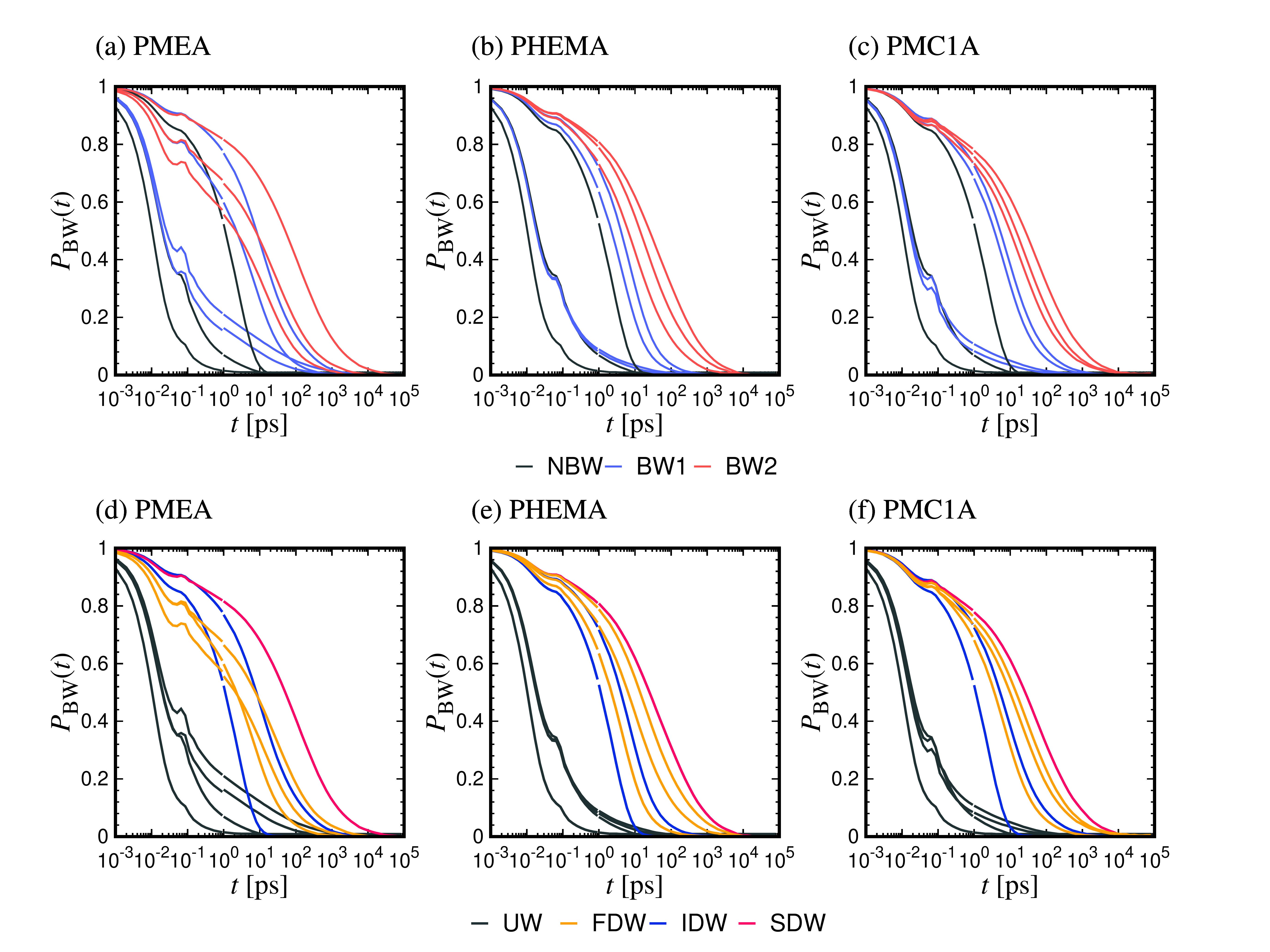}
\caption{Relaxation of 10 water molecule states characterized by the
 time correlation function $\pbw(t)$ at 90 wt\% in PMEA [(a) and (d)],
 PHEMA [(b) and (e)], and PMC1A [(c) and (f)].
(a)-(c) Classification by the number of H-bonds with polymeric oxygen,
 namely, NBW, BW1, and BW2 are represented by black, blue, and red
 color, respectively.
Since NBW, BW1, and BW2 represent the number of H-bonds with the
 polymer, they correspond to 3, 4, and 3 curves, respectively. 
(d)-(f) Classification by acceptor oxygen of water molecule, namely, UW,
 FDW, IDW, and SDW, are represented by black, orange, blue, and red
 color, respectively.
In total, the 10 curves are the same between (a) and (d), (b) and (e),
 and (c) and (f). 
The difference is the classification scheme
 represented by color codes.
}
\end{figure}

\begin{figure}[htbp]
     \centering
     \includegraphics[width=0.65\linewidth]{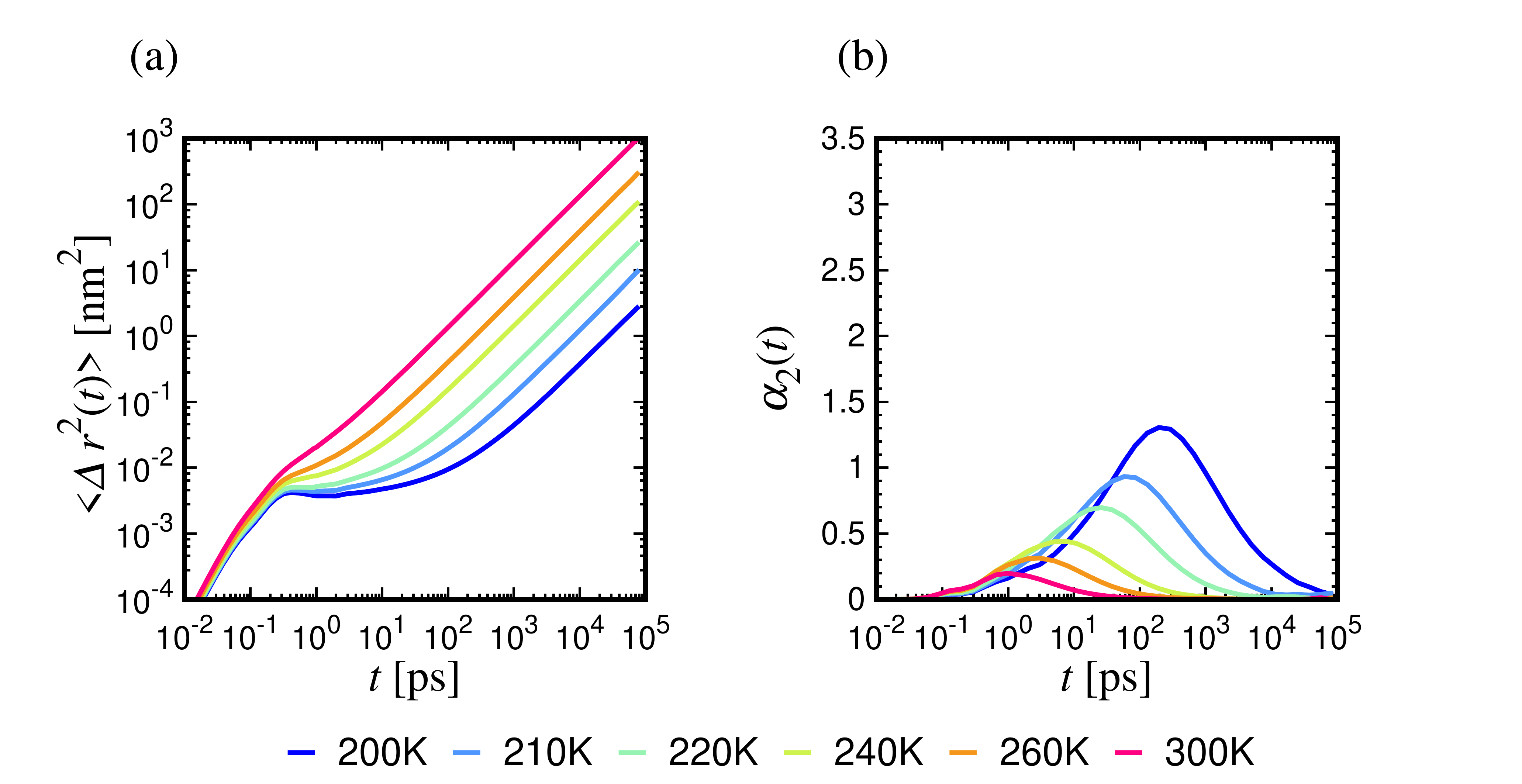}
     \caption{MSD $\langle\Delta r^2(t)\rangle$ (a) and NGP
 $\alpha_2(t)$ (b) 
in TIP4P/2005 liquid water system at 1 g/cm$^3$.}
\end{figure}

\begin{figure}[htbp]
\centering
\includegraphics[width=0.9\linewidth]{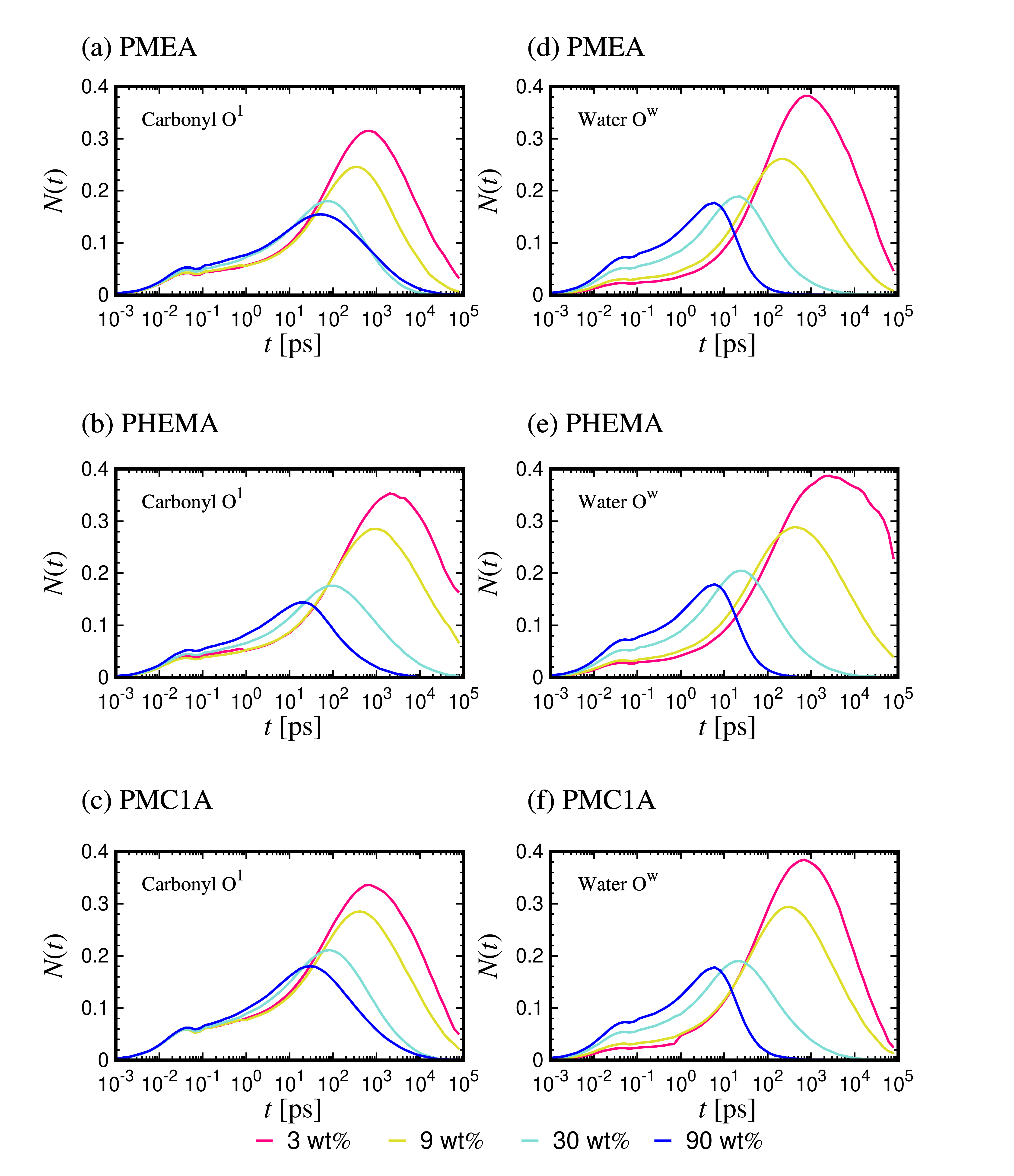}
\caption{Time correlation function $N(t)$ evaluating the contribution of
 the water molecule that does not diffuse after the breakage of H-bond
 with (a)-(c) carbonyl and (d)-(e) water in PMEA [(a) and (d)], PHEMA
 [(b) and (e)], and PMC1A [(c) and (e)].
}
\end{figure}

\begin{figure}[htbp]
\centering
\includegraphics[width=0.5\linewidth]{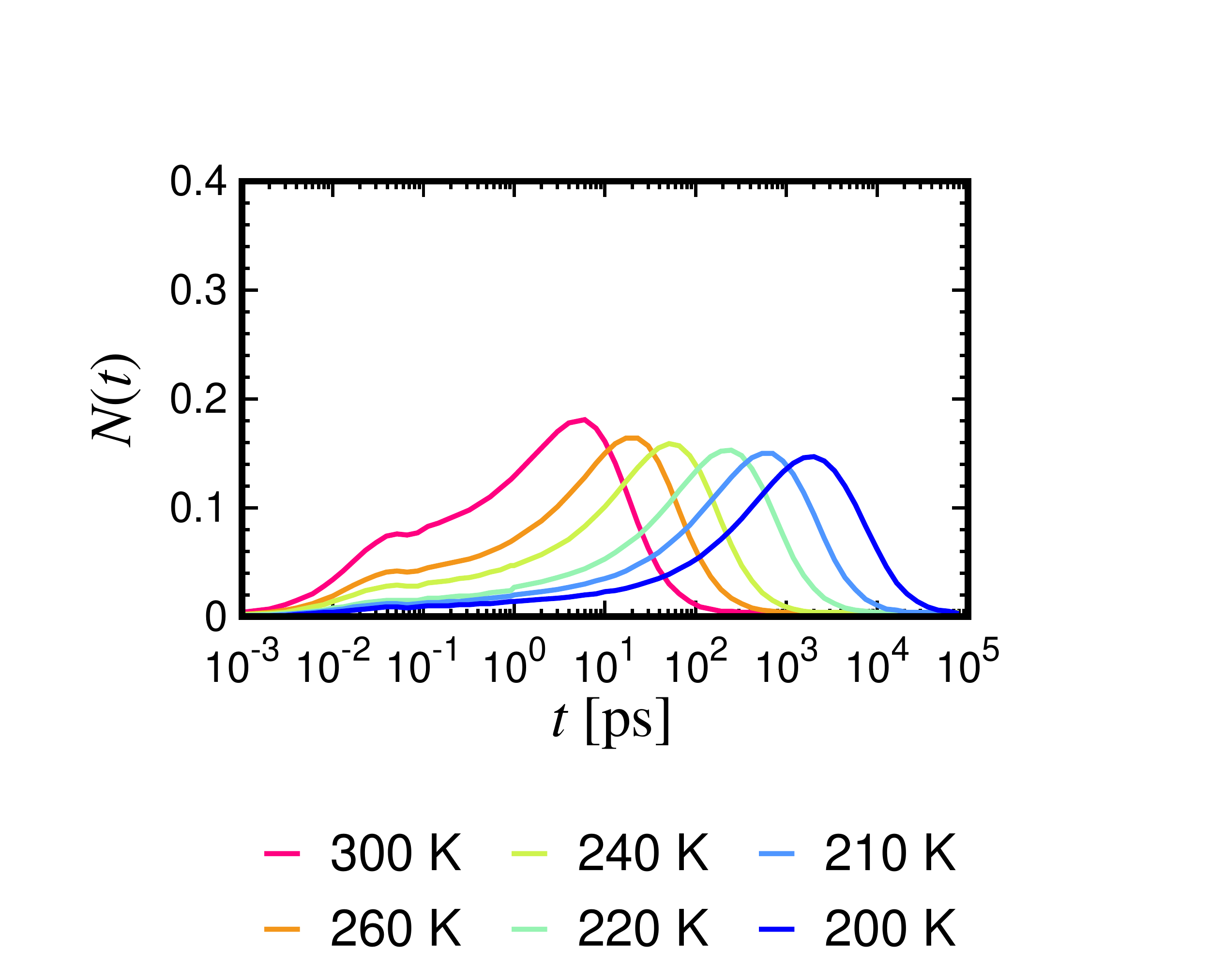}
\caption{Time correlation function $N(t)$ in bulk water of the
 TIP4P/2005 model at 1 g/cm$^3$.
The results demonstrate the dependence on temperature, with decreasing temperatures from left to right.}
\end{figure}

\end{document}